\documentclass[pdflatex,sn-mathphys-num]{sn-jnl}%
\usepackage{graphicx}%
\usepackage{multirow}%
\usepackage{amsmath,amssymb,amsfonts}%
\usepackage{amsthm}%
\usepackage{mathrsfs}%
\usepackage[title]{appendix}%
\usepackage{xcolor}%
\usepackage{textcomp}%
\usepackage{manyfoot}%
\usepackage{booktabs}%
\usepackage{algorithm}%
\usepackage{algorithmicx}%
\usepackage{algpseudocode}%
\usepackage{listings}%

\usepackage{bm}
\usepackage{color}

\newcommand{\meanrho}{\overline{\rho}}

{}
\newcommand{\meanUU}{\overline{\bm{U}}}
\newcommand{\meanVV}{\overline{\bm{V}}}
{}
{}

\newcommand{\meanN}{\overline{n}}

\newcommand{\meanT}{\overline{T}}

\newcommand{\meanU}{\overline{U}}
\newcommand{\meanS}{\overline{S}}

\begin{document}

\title{\bf Turbophoresis of inertial particles in inhomogeneous turbulence generated by oscillating grids}

\author{Elad~Elmakies}
\email{eladelm@post.bgu.ac.il}
\author{Oleg~Shildkrot}
\email{olegshil@post.bgu.ac.il}
\author{Nathan~Kleeorin}
\email{nat@bgu.ac.il}
\author{Avi~Levy}
\email{avi@bgu.ac.il}
\author{Igor~Rogachevskii}
\email{gary@bgu.ac.il}

\affil{The Pearlstone Center for Aeronautical Engineering
Studies, Department of Mechanical Engineering, The Ben-Gurion
University of the Negev, POB 653, Beer-Sheva 8410530, Israel}

%\date{Received: date / Revised version: date}

\abstract{
Turbophoresis in inhomogeneous turbulent flows leads to the formation of large-scale nonuniform particle number density distributions of inertial particles. This effect is associated with an effective drift velocity directed toward regions of lower turbulence intensity.
It depends on the Stokes and Reynolds numbers, as well as on the gradient of the turbulence intensity.
In the present study, turbophoretic transport is experimentally investigated  in air turbulent flows
generated by one-grid and two-grid oscillating systems.
The flow velocity field is measured using Particle Image Velocimetry, and the particle spatial distribution is obtained by applying image processing techniques.
To isolate the effect of particle accumulation due to turbophoresis from that associated with mean fluid  flow, the measured particle number density of inertial particles is normalized by the corresponding distribution obtained for noninertial tracer particles under identical flow conditions. The measurements show preferential accumulation of inertial particles in regions of minimum mean-square turbulent velocity, consistent with the expected behavior of turbophoretic transport.
}


\maketitle

\section{Introduction}
\label{sect1}

Transport of particles in turbulent flows has been studied extensively through theoretical analysis, laboratory and field experiments, and numerical simulations; see, e.g., Refs.~\cite{MY71,MY75,CSA80,MC90,ZRS90,F95,BLA97,P2000,SP06,LE08,CST11,DA13,RI21}. Nevertheless, several important aspects of particle transport in turbulence remain unresolved, especially in geophysical and astrophysical flows, where the governing parameters are difficult to reproduce experimentally or numerically.

A central problem in turbulent transport of inertial particles is the formation of small-scale and large-scale clusters
of inertial particles in turbulent flows (see, e.g., Refs.~\cite{ZA08,WA09,BE10,MBC12}).
Small-scale clusters are produced at scales much smaller than the integral turbulence scales, while
large-scale clusters (i.e., large-scale inhomogeneous particle number density distributions)
are formed at scales exceeding the integral turbulence scale.
Two key turbulence-related mechanisms can lead to large-scale particle accumulation.
The first mechanism is turbulent thermal diffusion in temperature-stratified turbulence, which causes inertial and noninertial particles to accumulate near regions of fluid mean temperature minimum \cite{RI21,EKR96,EKR97,EKRS00,EKRS01}.
The second mechanism of large-scale particle accumulation is turbophoresis, which results in
migration of inertial particles toward regions of lower turbulence intensity \citep{CTT75,RE83}.

Turbophoresis is due to the joint effect of particle inertia and spatial inhomogeneity of turbulence.
This phenomenon originates from averaging of particle velocity,
so that the mean particle velocity $\meanUU^{\rm(p)}$
is given by $\meanUU^{\rm(p)} = \meanUU + \meanVV_{\rm turboph} + {\bm V}_{\rm g}$,
where $\meanUU$ is the mean fluid velocity,
${\bm V}_{\rm g} = \tau_{\rm p} \, {\bm g}$
is the terminal fall velocity of inertial particles in the gravity field, and
\begin{eqnarray}
\overline{\bm V}_{\rm turboph} = - \kappa_{\rm turboph} \, {\bm \nabla} \left\langle{\bm u}^2\right\rangle
\label{G11}
\end{eqnarray}
is the turbophoretic velocity of inertial particles in an inhomogeneous turbulence,
$\tau_{\rm p}$ is the Stokes time
and $\kappa_{\rm turboph}$ is the turbophoretic coefficient, which depends on the Stokes and Reynolds numbers.
Here the quantities with overline denote mean fields and the angular brackets imply ensemble averaging.
The parameter $\kappa_{\rm turboph}$ vanishes for noninertial particles, so that
turbophoretic transport occurs only for inertial particles.

Turbophoresis has been studied extensively using analytical and numerical approaches in different types of turbulent flows
\citep{G97,EKR98,MS02,G08,SSB12,LCB16,MHR18,JBM20,BV24,WZZ24}.
Turbophoresis may be important for various geophysical, planetary and astrophysical inhomogeneous turbulent flows \cite{EKPR97,SSEKR09,H16,KR25}.
Direct numerical simulations (DNS) demonstrate that inertial particles in inhomogeneously forced isothermal turbulence accumulate preferentially in regions of minimum mean-square turbulent velocity \citep{MHR18}. In these flows, the spatial distribution of inertial particles is governed by the competition between turbophoretic transport and turbulent diffusion. The DNS results also show that the product $\kappa_{\rm turboph}$ and $u_{\rm rms} \equiv \sqrt{\langle{\bm u}^2\rangle}$ increases linearly with ${\rm St}$ for ${\rm St}\ll 1$, reaches a maximum near ${\rm St} \sim 10$, and decreases as ${\rm St}^{-1/3}$ for large ${\rm St}$. Here, ${\rm St}= \tau_{\rm p}/\tau_0$ is the Stokes number defined as a ratio of the Stokes time $\tau_{\rm p}$ and the characteristic turbulent time scale $\tau_0$ associated with the integral turbulence scale.
Numerical studies of turbophoresis in inhomogeneous turbulent Kolmogorov flows also demonstrate that the large-scale clustering of inertial particles increases at small ${\rm St}$ and gradually weakens at larger ${\rm St}$   \citep{LCB16}. Although these studies have not explicitly interpreted the results in terms of a balance between turbophoretic and turbulent diffusive fluxes, the observed particle accumulation reflects the combined action of turbophoretic and turbulent diffusive transport.

Despite substantial progress in theoretical and numerical studies, direct experimental investigations of turbophoresis remain relatively limited, particularly in nearly unbounded turbulent flows generated away from solid boundaries. Most previous experimental studies of particle accumulation have been conducted in wall-bounded turbulent flows, where turbulence inhomogeneity is strongly affected by walls and mean shear \cite{SE91,KHB95,RR04}. The experimental identification of turbophoretic transport is therefore challenging, especially in flows where multiple particle transport mechanisms may coexist.

In the present study, we experimentally investigate the formation of large-scale concentrations of inertial particles in strongly inhomogeneous isothermal turbulence generated by one and two oscillating grids in air flows. Oscillating-grid turbulence provides a controlled laboratory configuration with relatively weak mean flow and spatially varying turbulence intensity, allowing direct investigation of particle accumulation induced by turbophoresis. Complementary measurements of turbulent velocity fields and particle number density distributions allow direct comparison between the spatial distribution of inertial particles and the turbulence intensity.

The goal of this paper is to provide direct laboratory evidence of large-scale accumulation of inertial particles
in oscillating-grid turbulence and to test whether the measured inertial-particle distribution follows
the expected spatial trend relative to the turbulence-intensity field known
from analytical and numerical studies of turbophoresis.
The robustness of this trend is examined in two distinct turbulent-flow configurations generated
by one oscillating grid and by two oscillating grids.
These configurations differ in the spatial structure of turbulence intensity, anisotropy,
integral length scales, characteristic turbulent time, and Reynolds-number distribution.
To isolate the turbophoretic contribution, the measured particle number density fields
for inertial particles are normalized using corresponding measurements obtained
with noninertial particles under identical flow conditions.
The experiments show that in both turbulent-flow configurations, inertial particles preferentially
accumulate in regions of lower turbulence intensity.

This paper is organized as follows.
Section~\ref{sect2} discusses a characterisation of turbophoresis.
Section~\ref{sect3} describes the experimental setup, measurement techniques,
data analysis and the uncertainty and reliability of the measurements.
Section~\ref{sect4} analyzes the experimental results.
Finally, Section~\ref{sect5} discusses the obtained experimental results and
outlines the conclusions.

\section{Nature of turbophoresis and its characterisation}
\label{sect2}

The equation of motion for small inertial particles with the material density $\rho_{\rm p}$ that is much larger than the fluid density  $\rho$, is given by
\begin{eqnarray}
 \frac{{\rm d} {\bm U}^{\rm(p)}} {{\rm d} t} =- \frac{{\bm U}^{\rm(p)}-{\bm U}}{\tau_p} + {\bm g},
\label{G1}
\end{eqnarray}
where ${\bm U}^{\rm(p)}$ is the particle velocity, ${\bm U}$ is the fluid velocity,
${\bm g}$ is the acceleration caused by the gravity field, and
$\tau_{\rm p}=m_{\rm p} /(3 \pi \rho \, \nu \,d_{\rm p})$ is the Stokes
time of spherical solid particles having the diameter $d_{\rm p}$ and mass
$m_{\rm p} = \left(\pi /6\right)\, \rho_{\rm p}\, d_{\rm p}^3$, and $\nu$  is the kinematic viscosity
of the fluid.
Equation~(\ref{G1})  is valid for small inertial particles when the point-particle approximation and the Stokes drag approximation are satisfied.

The spatial distribution of inertial particles is determined by equation for the particle number density $n(t,{\bm x})$:
\begin{eqnarray}
\frac{\partial n}{\partial t} + {\bm \nabla} {\bf \cdot}\left({\bm U}^{\rm(p)}\, n\right) = D \, \Delta n ,
\label{G20}
\end{eqnarray}
where $D= k_B \,T/(3\pi \rho \, \nu \, d_{\rm p})$ is the coefficient
of the molecular (Brownian) diffusion of particles, $T$ is the fluid temperature
and $k_B$ is the Boltzmann constant.
Averaging Eq.~(\ref{G20}) over an ensemble yields an equation for the mean particle number density $\meanN$:
\begin{eqnarray}
\frac{\partial \meanN}{\partial t} + {\bm \nabla} {\bf \cdot}\left[\meanUU^{\rm(p)} \meanN + \langle {\bm u}^{\rm(p)}  \, n'  \rangle \right]  = D \, \Delta \meanN ,
\label{G12}
\end{eqnarray}
where ${\bm u}^{\rm(p)}$ are fluctuations of particle velocity and $n'$ are particle number density fluctuations.
Mean-field equation~(\ref{G12}) for inertial particles has been derived
by means of various analytical approaches (see, e.g., Refs.~\cite{ZRS90,ZA08,RI21}).

For large P\'eclet numbers, the molecular Brownian diffusion coefficient $D$  can be neglected,
because it is much smaller than turbulent diffusion coefficient.
In Eq.~(\ref{G12}), the mean particle velocity is
$\meanUU^{\rm(p)} = \meanUU + \tau_{\rm p} \, {\bm g} - \kappa_{\rm turboph} \, {\bm \nabla} \left\langle{\bm u}^2\right\rangle$
and we take into account that the turbulent flux of particles $\langle u^{\rm(p)}_i  \, n'  \rangle$ in an anisotropic
turbulence is given by $\langle u^{\rm(p)}_i  \, n'  \rangle = \left\langle u_i  \, n'  \right\rangle + {\rm O}(\tau_{\rm p}/\tau_0)
= - D_{ij}^{^{\rm (T)}} \nabla_j \meanN$, where $\tau_0$ is the characteristic turbulent time scale associated
with the integral turbulence scale and $D^{^{\rm (T)}}_{ij}$ is the turbulent diffusion tensor.
In the turbulent flux of particles we neglect small effects $\sim {\rm O}(\tau_{\rm p}/\tau_0)$.

For simplicity, we assume that gradients of $\left\langle{\bm u}^2\right\rangle$ and $\meanN$
along the horizontal direction (e.g., along the $Y$ axis)
dominate over gradients in the remaining directions, consistent with the experimental
conditions considered in the present study (see Sect.~\ref{sect3}).
The steady-state solution of Eq.~(\ref{G12}) for inertial particles at a zero total particle flux at the boundaries
is given by
\begin{eqnarray}
\frac{\meanN_{\rm in}}{\meanN_0} =   \exp \biggl(\int_{Y_0}^{Y}
\frac{\meanU_y - \kappa_{\rm turboph} \, \nabla_y \left\langle{\bm u}^2\right\rangle}{ D^{^{\rm (T)}}_{yy} } \,{\rm d} Y' \biggr) ,
\label{G14}
\end{eqnarray}
where $D^{^{\rm (T)}}_{yy}=\ell_y \, \left[\left\langle u_y^2\right\rangle\right]^{1/2}
= \tau_0 \left\langle {\bm u}^2\right\rangle/a_\ast$ is the horizontal component of the turbulent diffusion tensor,
$\ell_y$ is the integral scale along the $Y$ axis,
$a_\ast$ is a coefficient that depends on turbulence anisotropy
(e.g., $a_\ast =3$ for isotropic turbulence and $a_\ast>3$ for anisotropic turbulence),
$\meanN_{\rm in}$ is the mean number density of inertial particles
and $\meanN_0 = \meanN_{\rm in}(Y=Y_0)$.
Equation~(\ref{G14}) can be rewritten as
\begin{eqnarray}
\frac{\meanN_{\rm in}}{\meanN_0} =  \exp \biggl(\int_{Y_0}^{Y} \frac{\meanU_y}{ D^{^{\rm (T)}}_{yy} } \,{\rm d} Y' \biggr)
\, \left(\frac{\left\langle{\bm u}^2\right\rangle(Y)}{\left\langle{\bm u}^2\right\rangle_0}\right)
^{- \frac{a_\ast \kappa_{\rm turboph}}{\tau_0} } ,
\label{GG14}
\end{eqnarray}
where $\left\langle{\bm u}^2\right\rangle_0 =\left\langle{\bm u}^2\right\rangle(Y=Y_0)$, and
we assume that the parameters $\kappa_{\rm turboph}$ and $\tau_0$ vary weakly along the horizontal coordinate $Y$.
For noninertial particles the turbophoretic velocity vanishes, and Eq.~(\ref{G14}) reads
\begin{eqnarray}
\frac{\meanN_{\rm nin}}{\meanN_0} =  \exp \biggl(\int_{Y_0}^{Y} \frac{\meanU_y}{ D^{^{\rm (T)}}_{yy} } \,{\rm d} Y' \biggr) ,
\label{GG15}
\end{eqnarray}
where  $\meanN_{\rm nin}$ is the mean number density of noninertial particles
and $\meanN_0 = \meanN_{\rm nin}(Y=Y_0)$.
Using Eqs.~(\ref{GG14})--(\ref{GG15}), we obtain that the ratio of the mean number densities of inertial particles $\meanN_{\rm in}$
and noninertial particles $ \meanN_{\rm nin}$ for identical flow conditions as
\begin{eqnarray}
\frac{\meanN_{\rm in}(Y)}{\meanN_{\rm nin}} \approx \left(\frac{\left\langle{\bm u}^2\right\rangle(Y)
}{\left\langle{\bm u}^2\right\rangle_{\rm max}}\right)^{- \frac{a_\ast \kappa_{\rm turboph}}{\tau_0}} ,
\label{G15}
\end{eqnarray}
where $\left\langle{\bm u}^2\right\rangle_{\rm max}$ is the maximum turbulence intensity in the flow domain
at $Y=Y_0$. It follows from Eq.~(\ref{G15}) that the ratio $\meanN_{\rm in}/\meanN_{\rm nin}$ does not explicitly depend on the mean
flow velocity $\meanU_y$. We will use this fact in the data analysis of the experiments on turbophoresis,
by normalising the mean number density of inertial particles $\meanN_{\rm in}$ by that for noninertial particles $\meanN_{\rm nin}$.
Equation~(\ref{G15}) also implies that inertial particles are accumulated in the regions of low turbulence intensity.

The physical mechanism underlying turbophoresis is associated with particle inertia in spatially inhomogeneous turbulence. Due to centrifugal effects, inertial particles cannot fully follow fluid turbulent motions and tend to drift out to the regions between turbulent eddies where fluid velocity fluctuations vanish and fluid pressure fluctuations are maximum. In homogeneous and isotropic turbulence without preferred direction, this drift does not produce a net large-scale particle transport. However, in the presence of a nonzero turbulence intensity gradient $({\bm \nabla}\left\langle{\bm u}^2\right\rangle \not= {\bm 0})$, the particle drift results in a net turbophoretic flux, and inertial particles preferentially migrate to the regions with reduced turbulence intensity.

It is important to distinguish between phenomena of turbophoresis and turbulent thermal diffusion, since these mechanisms originate from different physical nature. Turbulent thermal diffusion produces an additional nondiffusive turbulent flux $\langle {\bm u}^{\rm(p)}  \, n'  \rangle$ of particles ${\bm J}_{\rm TTD}  = \meanN \, {\bm V}_{\rm TTD}$ \cite{RI21,EKR96,EKR97}, where
${\bm V}_{\rm TTD}=- \alpha \, D_{\rm T} \, {\bm \nabla} \ln \meanT$
is the effective pumping velocity in a temperature-stratified turbulence, $\meanT$ is the mean fluid temperature,
$D_{\rm T}$ is the turbulent diffusion coefficient, and the parameter $\alpha$
depends on the Stokes and Reynolds numbers for inertial particles, while $\alpha=1$ for noninertial particles.
Turbulent thermal diffusion is a collective phenomenon, it occurs in temperature-stratified turbulence and is associated with turbulent particle flux (correlation of particle velocity and number density fluctuations) coupled to turbulent heat flux
(correlation of fluid velocity and temperature fluctuations).
This phenomenon has been investigated theoretically, numerically, in laboratory experiments
as well as in geophysical, planetary and astrophysical turbulence applications \cite{EKPR97,SSEKR09,H16,KR25,AEKR17,HKRB12,RKB18}.
In contrast, turbophoresis originates from averaging of particle velocity \citep{CTT75,RE83}, and it is
not related to the correlation of particle velocity and number density fluctuations.

\section{Experimental setup and measurement technique}
\label{sect3}

In this section we describe the experimental setup, measurement techniques,
data analysis, as well as the uncertainty and reliability of the measurements.

\subsection{Experimental setup}

\begin{figure*}[t!]
\centering
\includegraphics[width=6.0cm]{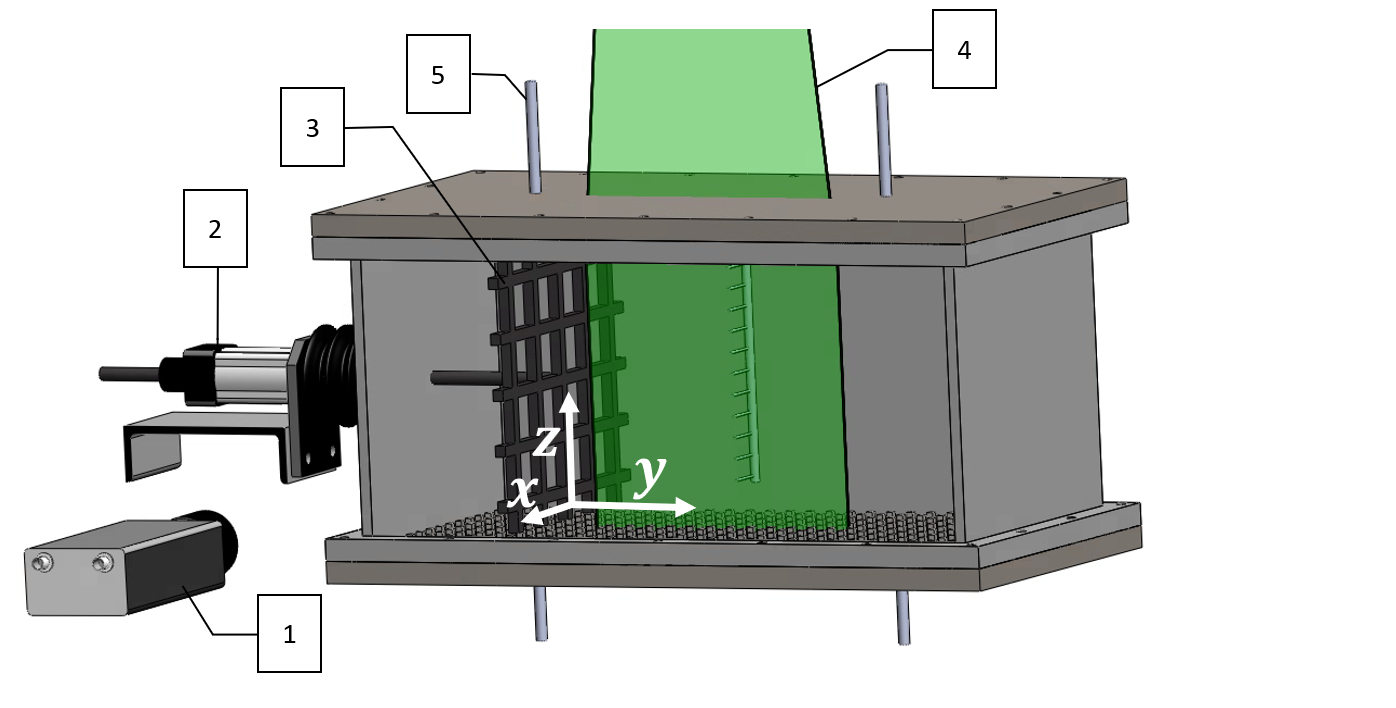}
\includegraphics[width=6.0cm]{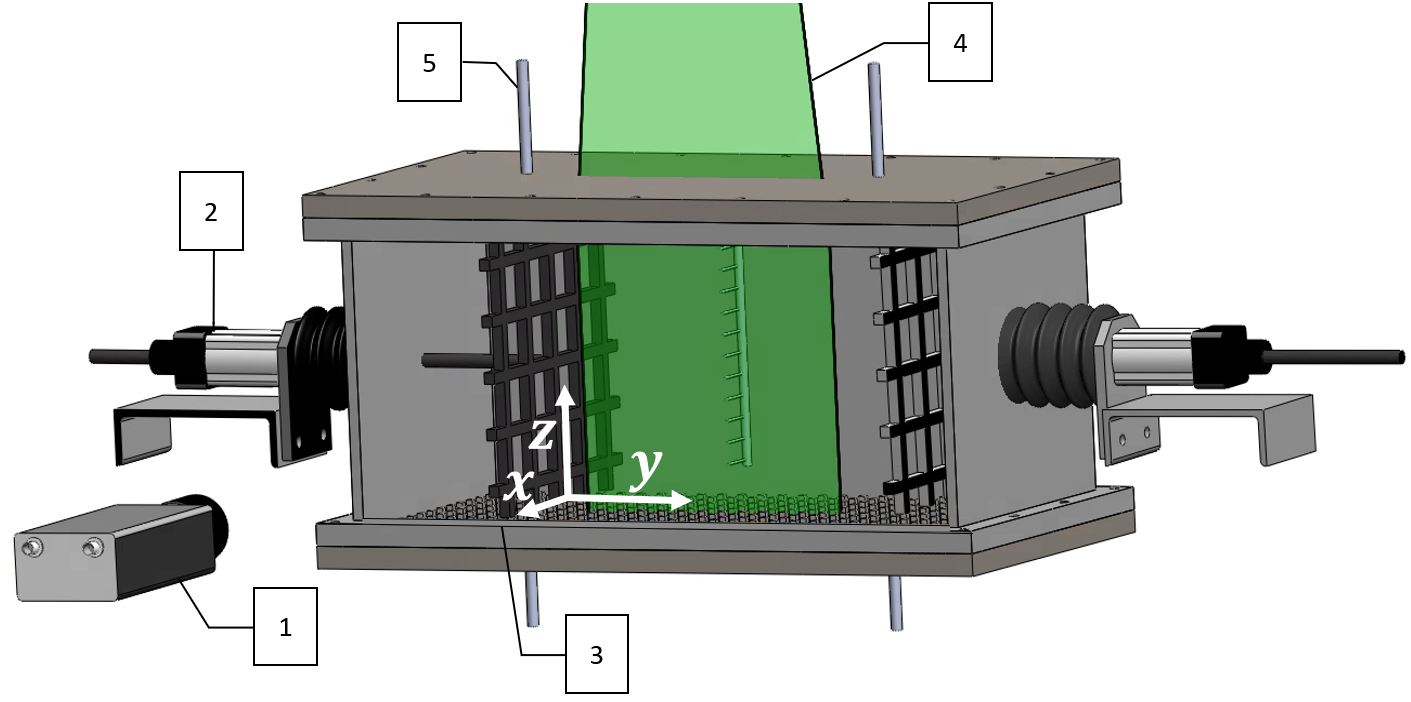}
\caption{\label{Fig1}
Experimental setup  with an isothermal turbulence
produced by one oscillating grid  (left panel) and by two oscillating grids (right panel): (1) digital CCD camera; (2) rod driven by the
speed-controlled motor; (3) oscillating grid;  (4) laser light sheet;
(5) the inlets of acoustic feeding device for particle injection.
}
\end{figure*}

Experiments on turbophoresis of inertial particles are performed in isothermal inhomogeneous turbulence generated by one and two oscillating grids in air flows.
The experiments are conducted in a rectangular transparent chamber with dimensions $L_x \times L_y \times L_z$, where $L_x=L_z=26$ cm and $L_y=53$ cm. The $Z$ axis is vertical, while the $Y$ axis is perpendicular to the grid plane  (see Fig.~\ref{Fig1}).

Each oscillating grid has dimensions of $25\times25$ cm$^2$ and consists of a square array of bars with a mesh size of 5 cm
and a bar width of 1 cm. The grids are positioned in the $XZ$ plane, parallel to the chamber side walls at a distance of 10 cm, corresponding to two grid meshes from the side walls of the chamber.
The grids oscillate along the $Y$ direction with frequency $f=10.5$  Hz and stroke amplitude $A_g=6$ cm.
The grid Reynolds number based on the grid mesh size $M=5$ cm is Re$_{\rm M}=f M^2/\nu=1750$.
The grids drive the carrier flow and generate the inhomogeneous turbulence.

Each grid performs an approximately sinusoidal translational motion normal to the grid plane.
The motion is described as simple harmonic motion.
In the double-grid configuration, the two grids are driven by separate motors.
They are operated at the same nominal frequency and stroke amplitude,
but they are not mechanically or electronically connected.
Therefore, no fixed phase difference between the two grids is prescribed during the measurements.
The relative phase is not controlled and is allowed to vary during the acquisition.
The values of the frequency and amplitude are chosen as fixed operating conditions
that provide sufficiently strong and reproducible turbulence in the measurement region,
while maintaining stable grid motion, adequate optical access, and reliable velocity
and particle-number-density measurements.

Rectangular pins of size $3 \times 3 \times 15$ mm are mounted on the upper and lower walls
of the chamber (see Fig.~\ref{Fig1}) to reduce the mean flow velocity in the core region and
to enhance the spatial inhomogeneity of the mean flow.
The rectangular pins are passive flow-modifying elements.
They do not generate the turbulence that is produced by the oscillating grids.
Their role is to disturb the residual large-scale circulation and introduce local blockage
and wake regions near the boundaries, thereby enhancing the spatial gradients
of turbulence intensity in the measurement region.
The resulting inhomogeneity is quantified directly from the velocity measurements.

\subsection{Measurement technique and data analysis of velocity field}

The velocity field is measured using a Particle Image Velocimetry (PIV) system \cite{AD91,RWK07,W00} consisting of a Nd-YAG laser (Continuum Surelite, $2 \times 170$ mJ) with a wavelength of 532 nm and a 12-bit CCD camera.
The velocity fields are measured in a flow domain of $209.1 \times 155.4$ mm$^2$ using images with a resolution of $1376 \times 1024$ pixels, corresponding to a spatial resolution of 151 $\mu$m/pixel. The velocity field in the probed region is analyzed using interrogation windows of $16 \times 16$ pixels with 25 \% overlap, and the pulse separation is $\Delta t=1500 \,\mu$s.
The laser light sheet of the PIV system is located in the central vertical $YZ$ plane at $X=0$.
Thus, the velocity measurements are performed in the central optical plane of the chamber.
The laser provides a single-pulse energy of 170 mJ.
The PIV image pairs are acquired at the rate 2 Hz, and the laser is externally triggered at the same pulse-pair acquisition rate.
The acquisition is not synchronized with the oscillating grid motion, whose frequency is 10.5 Hz.

The velocity statistics are obtained from 530 image pairs acquired over many grid-oscillation cycles rather than at a fixed phase of the grid motion.
Incense smoke with solid spherical particles is used as a tracer for the PIV measurements.
The particles have a mean diameter of $0.7 \,\mu$m and a material density $\rho_{\rm p}\approx 10^3 \meanrho$,
where $\overline{\rho} = 10^{-3}$ g cm$^{-3}$ is the mean fluid density.
The particles are generated by high-temperature sublimation of solid incense grains (see Ref.~\cite{EKRL22} for details).

Using the velocity measurements, we determine various mean-flow and turbulent characteristics; see Sec.~\ref{sect4} for details. In particular, the mean velocity field, mean velocity shear, root-mean-square (r.m.s.) velocities, turbulent anisotropy parameter, two-point velocity correlation functions and integral turbulence length scales are obtained. The mean and r.m.s. velocity fields are determined at each spatial location by averaging over 530 independent velocity maps.

The mean velocity and rms velocity profiles are also recalculated using subsets of the recorded data, and they are compared with the final profiles obtained from all 530 velocity maps. The profiles obtained from the later subsets and from the full data set showed only small variations, indicating that the statistical estimates are stable with respect to the number of velocity maps used in the averaging.
The profiles changed only very weakly when the number of velocity maps included in the averaging is varied.

The consecutive velocity maps are separated by 0.5 s (since they are acquired at the rate 2 Hz).
This timescale is much larger than the characteristic turbulent correlation time $\tau_0$ in the measurement region
(e.g., $\tau_0 \approx 0.05$ s is for one-grid configuration, and $\tau_0 \approx 0.03$ s is for two-grid configuration).
Therefore, the velocity maps used in the averaging can be treated as statistically independent for the purpose of estimating the mean and rms velocity fields.
In addition, the size of the probed region does not affect the obtained results.

The integral  length scales of the turbulence $\ell_y$ and $\ell_z$ in the horizontal $Y$ and vertical $Z$ directions are determined from two-point velocity correlation functions obtained in many locations. Two standard approaches are applied and yield similar results. In the first approach, the normalized two-point correlation function of the velocity fluctuations is integrated over the separation distance $r$ up to the point where the correlation decays to zero. In the second approach, the integral length scale is determined from an exponential fit of the form $\exp(-r^2/\ell_0^2)$ to the normalized correlation function.

\subsection{Measurement technique and data analysis of inertial particle distributions}

\begin{table}
\caption{\label{table-1}
Characteristics of inertial particles}
\begin{tabular}{| c| c|}

\hline
Particle mean diameter ($\mu$m)               &  10                            \\
\hline
Particle diameter range   ($\mu$m)            &  2 - 20                        \\
\hline
Particle material density (g/cm$^3$)        & 1.1                                \\
\hline
Mass loading parameter, $n \, m_p/\rho$  & $5.8 \times 10^{-4}$  \\
\hline
Volume fraction                                      & $5.2 \times 10^{-7}$  \\
\hline
Settling velocity (cm/s)                          &  0.33                              \\
\hline
\end{tabular}
\end{table}

For experimental study of turbophoresis, borosilicate hollow glass particles with an approximately spherical shape are used as inertial particles. The inertial particles do not affect the fluid flow because the mass-loading parameter is small ($m_{\rm p} n / \rho \ll 1$).
Characteristics of inertial particles are given in Table~\ref{table-1}.
In the experiments with inertial particles,
the Stokes number ${\rm St}= \tau_{\rm p}/\tau_0$ based on the Stokes time $\tau_{\rm p}$ and the characteristic turbulent time scale $\tau_0$ associated with the integral turbulence scale varies in the range $(5.3 - 8.2) \times 10^{-3}$ for turbulence produced by one oscillating grid and $(8.0 -12.3) \times 10^{-3}$ for turbulence produced by two oscillating grids
(see Table~\ref{table-2} in Section~\ref{subsect4.1}).
On the other hand,  the Stokes number ${\rm St}$ for noninertial particles having the diameter 0.7 $\mu$m, is about
$3.6 \times 10^{-5}$ for the one-grid configuration and $5.5 \times 10^{-5}$ for the two-grid configuration.

The spatial distributions of particles with diameters of $0.7\,\mu$m and $10\,\mu$m are obtained in separate experiments using Mie light scattering measurements \cite{guib01}. The mean scattered-light intensity is determined using interrogation windows of $32 \times 32$ pixels with 50\% overlap. For turbulence generated by one oscillating grid, the measurements are performed on a grid of $85 \times 64$ interrogation windows, while for turbulence generated by two oscillating grids the measurements are performed on a grid of $65 \times 51$ interrogation windows. This procedure enables determination of the distributions of scattered-light intensity in vertical strips composed of interrogation windows.

The scattered-light intensity produced by small particles is determined by Mie scattering \cite{BH83}. The scattered radiation energy flux $E_s$ depends on the particle diameter $d_{\rm p}$, the wavelength $\lambda$, and the refractive index through the scattering function $\Psi(\pi d_{\rm p}/\lambda;a_0,n)$. For $\lambda \gtrsim \pi d_p$, the scattering function follows the Rayleigh scattering regime and depends on particle diameter, whereas for $\lambda \lesssim \pi d_p$ the scattering function becomes weakly dependent on both particle diameter and wavelength. The scattered radiation energy flux $E_s$ is proportional to the local particle number density $n$. Therefore, the scattered-light intensity recorded by the CCD camera can be used to determine the spatial distributions of particle number density.

For a fixed particle type and optical configuration, the frame-averaged scattered-light intensity is proportional to the local particle number density. Therefore, the analysis is based on normalized relative particle-density fields rather than on absolute particle concentrations.
To isolate turbophoretic particle accumulation from other possible transport mechanisms, the scattered-light intensity $E_{\rm in}$ measured for inertial particles is normalized pointwise by the scattered-light intensity $E_{\rm nin}$ obtained in separate experiments with noninertial particles under identical flow conditions (see below). This normalization enables characterization of the spatial distribution of particle number density associated with turbophoretic transport.

The particles are introduced into the experimental chamber using a custom-made acoustic feeding
device designed to disperse the particles and reduce agglomeration.
The device consists of an acrylic chamber equipped with air guides and an acoustically driven flexible membrane that disperses particles into the incoming airflow.
To achieve approximately uniform particle dispersion within the flow domain,
the particles are injected into the chamber through two inlets located on the upper wall and two inlets positioned
on the lower wall of the chamber (see Fig.~\ref{Fig1}).
The inlets are small tubes used only to introduce particles into the chamber. They are not turbulence-generating elements.
Their diameter is small compared with the chamber height, and they are located near the walls,
away from the central optical measurement region.

Following the initial injection, the particles are allowed to disperse throughout the chamber
for approximately 60 seconds before Mie-scattering measurements are initiated,
so that the analyzed particle-number-density fields do not represent the immediate inlet region.
Since the inertial particles are not injected directly onto the grids, the analyzed
particle-number-density fields are obtained away from the immediate vicinity of the grid bars.
Therefore, the measured large-scale particle distributions
are interpreted as resulting from transport by the turbulent carrier flow
rather than from direct mechanical interaction with the oscillating grids.

For each particle type, the recorded images are divided into interrogation windows
of $32 \times 32$ pixels with 50 \% overlap.
The scattered-light intensity is spatially averaged within each interrogation window
and subsequently averaged over the recorded images to obtain the mean intensity distribution.
The analysis of the particle-number-density distributions involves two successive normalization steps.
First, for each particle type, the scattered-light intensity in each interrogation window is normalized
by the corresponding horizontal mean intensity at the same vertical coordinate $Z$.
This procedure serves three purposes:
\begin{itemize}
\item{
emphasizing particle redistribution along the horizontal $Y$ direction,
which is the principal direction of the turbulence-intensity gradient (see Sect.~\ref{sect4});}
\item{
reducing the influence of vertical variations in the measured scattered-light intensity,
including those associated with nonuniform laser illumination;}
\item{
establishing a common dimensionless reference for comparing particle
distributions obtained in separate experiments.}
The latter is particularly important because the absolute particle number density may differ between experimental runs, depending on the amount of injected particles and their dispersion within the chamber.
Normalizing each horizontal profile by its mean intensity removes differences in the overall intensity level and allows comparison of the relative spatial distributions independently of the absolute particle concentration.
\end{itemize}

Second, the horizontally normalized intensity distribution of inertial particles is divided pointwise by the corresponding normalized distribution of noninertial particles measured under identical flow conditions.
This additional normalization reduces the influence of spatial variations common to both particle populations,
including residual illumination nonuniformity, optical effects, and particle redistribution associated with the mean fluid flow.
The resulting dimensionless intensity ratio characterizes the preferential spatial redistribution
of inertial particles relative to noninertial tracers. The procedure therefore enables comparison
between separate experiments without requiring identical absolute particle concentrations or scattering intensities.
The normalization procedure is used to isolate turbophoretic particle accumulation from other transport effects.

To evaluate the statistical stability and repeatability of the particle-number-density measurements,
approximately five independent experimental runs have been performed for each investigated condition,
with approximately 500 images acquired in each run.
The measurements are initially analyzed separately in consecutive groups of 100 images.
The resulting normalized mean particle-density distributions are compared with those calculated
using the complete set of 500 images from each experimental run.
Consistent spatial distributions are obtained, indicating that the observed particle accumulation patterns
remain statistically stable over the measurement period.
Furthermore, the mean particle-density distributions obtained from the independent experimental runs were compared and showed consistent spatial patterns, demonstrating good experimental repeatability.
Following these comparisons, the instantaneous intensity maps from the independent runs
were combined to improve the statistical reliability of the final results.

It should be emphasized that the observed preferential accumulation of inertial particles refers to their nonuniform spatial distribution within the experimental chamber. The consistency of the normalized distributions over the measurement period and between independent experimental runs supports the establishment of an approximately statistically stationary spatial distribution under the investigated conditions.
Nevertheless, the present analysis does not independently quantify possible slow variations in the total particle concentration resulting from gravitational settling or deposition on the chamber walls.

The intensity-based method is selected because the primary objective of the present paper
is to study large-scale clustering of inertial particles and
to determine the time-averaged spatial distribution of particles over the entire measurement region,
rather than to identify or track individual particles.
An important experimental limitation is the spatial resolution of the imaging system.
The recorded images have a resolution of $1376 \times 1024$ pixels,
corresponding to approximately 151 $\mu$m/pixel in the measurement plane,
whereas the mean particle diameters are 10 $\mu$m for inertial particles and 0.7 $\mu$m for noninertial tracer particles.
Under the present imaging conditions, the effective spatial resolution and image quality
are insufficient for reliable identification, separation, and counting of individual particles throughout the measurement region.
Although individual particles may produce detectable scattered-light signals,
this does not necessarily allow each particle to be identified and counted independently.

Furthermore, direct particle counting would require reliable segmentation of individual particle images,
which introduces additional uncertainties associated with the different scattering efficiencies
of the two particle populations, variations in particle image brightness,
optical resolution, and the selection of detection thresholds.
In contrast, the intensity-based approach does not require individual particle identification.
Instead, the scattered-light intensity is spatially averaged within interrogation windows of $32 \times 32$
pixels and subsequently processed to determine the relative particle-density distribution.

For a given particle population under fixed optical conditions, the mean scattered-light intensity
provides a measure of the local particle number density, provided that the scattering properties
of the particle population remain statistically uniform throughout the measurement region.
As explained above in this section, the analysis involves two successive normalization steps.
The first establishes a common dimensionless reference for comparing particle distributions
obtained in separate experiments, while the second compares the normalized inertial-particle distribution
with the corresponding noninertial tracer distribution.
This approach enables characterization of large-scale preferential particle accumulation without
requiring individual particle counting or absolute calibration of particle concentrations.

The experimental methodology and data-processing procedures employed in the present study are based on approaches previously developed for investigations of
turbulent thermal diffusion \cite{BEE04,EEKR04,EEKR06a,EEKR06b,AEKR17,ZEKRL25,EKRL26},
turbulent convection \cite{BEKR09,EEKR11,BELR20,SKRL22,EKRL23}, stably stratified turbulence \cite{EEKR13,CEKR14}, particle mixing in inhomogeneous turbulence \cite{EHSR09,EKRL22}, and small-scale particle clustering \cite{EKR10}.

\subsection{The uncertainty and reliability of the measurements}

The uncertainty of the velocity measurements is estimated based on the accuracy of particle-image displacement
determination using cross-correlation techniques.
Previous studies \cite{AD91,RWK07,W00} demonstrate that subpixel
displacement accuracy can be achieved under suitable imaging conditions.
Assuming a characteristic displacement uncertainty of approximately 0.1 - 0.2 pixels,
together with the spatial calibration of 151 $\mu$m/pixel and a time separation of 1500 $\mu$s
between laser pulses, the corresponding velocity uncertainty is estimated to be approximately
$1 - 2$ cm/s for each measured velocity component.
This estimate represents the uncertainty associated with particle-image displacement determination.
The corresponding relative uncertainty depends on the local velocity magnitude and therefore varies
throughout the measurement region.
The mean velocity and turbulence statistics are obtained from 530 PIV velocity fields,
providing a statistical basis for determining the spatial distributions of the mean and fluctuating velocity components.

The reliability of the measurements is assessed by considering the main sources of uncertainty associated with the velocity and particle number density fields. The velocity field is measured using the PIV system, and its quality is evaluated using the standard $Q$-factor criterion, defined as the ratio between the primary and secondary peaks of the cross-correlation function \cite{AD91,RWK07,W00}. The spatially averaged $Q$ factor is approximately 1.6 or higher throughout the data set, indicating reliable cross-correlation and sufficient signal-to-noise ratio for accurate velocity measurements. Standard validation procedures are applied to remove spurious vectors and ensure robustness of the velocity field.

The statistical uncertainty of the particle-number-density measurements is evaluated
using first-order error propagation (the delta method).
The main statistical contributions originate from the frame-to-frame fluctuations of the scattered-light intensity
measured in the interrogation windows.
For each particle population, the uncertainty of the mean scattered-light intensity is estimated
using the standard deviation of the recorded intensity fluctuations
and the number of images used to calculate the corresponding mean value.

The particle-density analysis involves two successive normalization steps.
First, each intensity distribution is normalized by its horizontal mean at the corresponding vertical position.
This establishes a common dimensionless reference and reduces differences in the overall particle concentration
between experimental runs.
Second, the normalized inertial-particle distribution is divided pointwise by the corresponding
normalized noninertial-particle distribution.
The particle number density is determined from the ratio of frame-averaged scattered-light intensities measured for inertial and noninertial particles.
The $\delta$  method is used to estimate the statistical uncertainty associated with the measured intensity ratio,
considering the independent statistical contributions of the inertial particles measurements and noninertial
particles measurements.
In particular, the corresponding uncertainty is evaluated using first-order error propagation (the $\delta$ method), accounting for the independent contributions of both measurements \cite{BR03}, and varies from $4\%$ to $8\%$ with 95\% confidence interval.
Overall, the uncertainties in the velocity and particle number density measurements remain small relative to the observed flow variations and do not affect the main conclusions of the study.

To improve the reliability of the particle-number-density measurements, approximately five independent
experimental runs are conducted, each comprising approximately 500 images.
Each experimental run is initially analyzed separately, and consistent mean spatial distributions
are obtained under the same experimental conditions.
Following this verification, the instantaneous intensity distributions are combined for the final statistical analysis.
This procedure enabled us to assess the repeatability of the measurements before combining the experimental data.

\section{Experimental results}
\label{sect4}

This section presents the experimental results on large-scale inhomogeneous distributions of inertial particles in turbulence generated by one and two oscillating grids in airflow (see Fig.~\ref{Fig1}). The main objective is to investigate turbophoretic transport of inertial particles in strongly inhomogeneous turbulence at sufficiently large Reynolds numbers. Therefore, the turbulence characteristics are analyzed primarily to characterize the flow inhomogeneity responsible for turbophoretic transport, while the main focus is placed on the spatial distributions and accumulation of inertial particles.

\begin{figure*}[t!]
\centering
\includegraphics[width=6.0cm]{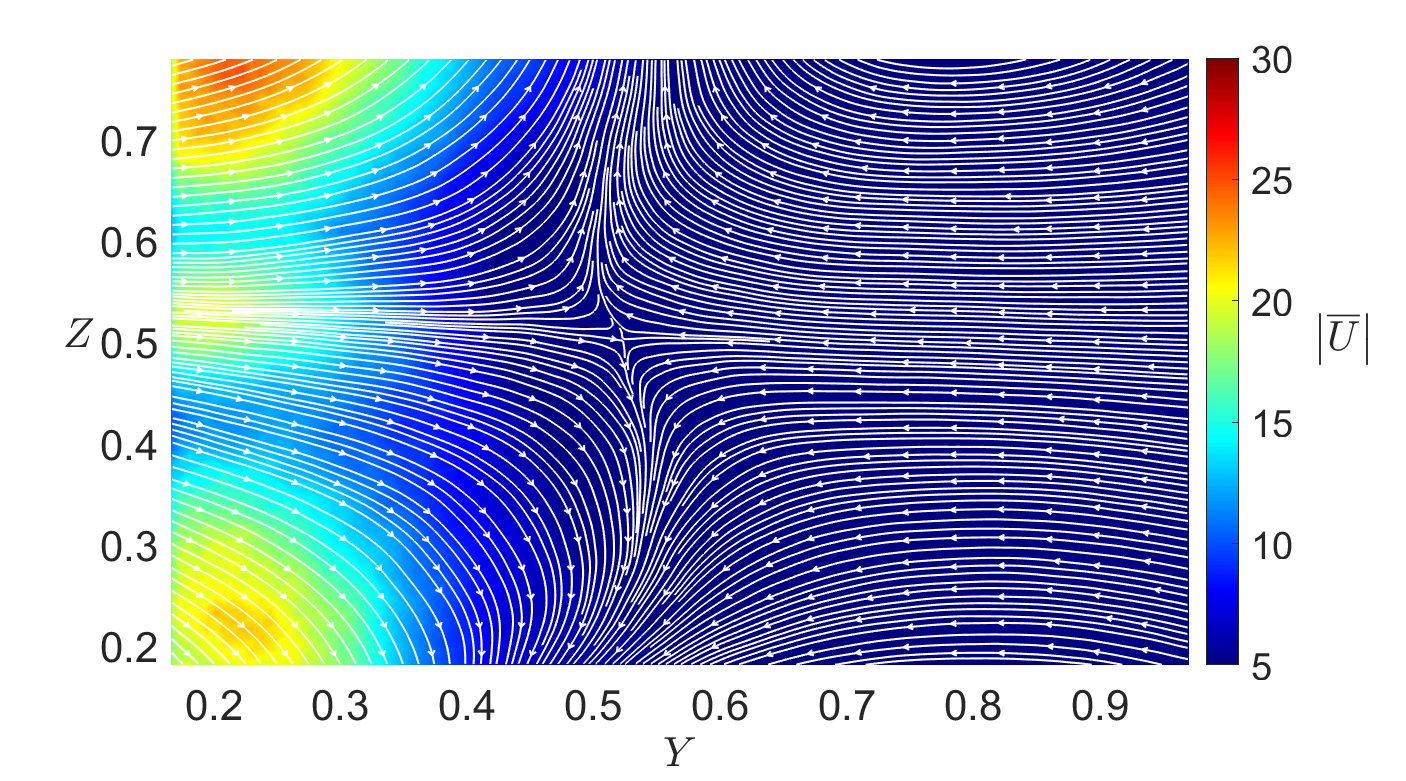}
\includegraphics[width=6.0cm]{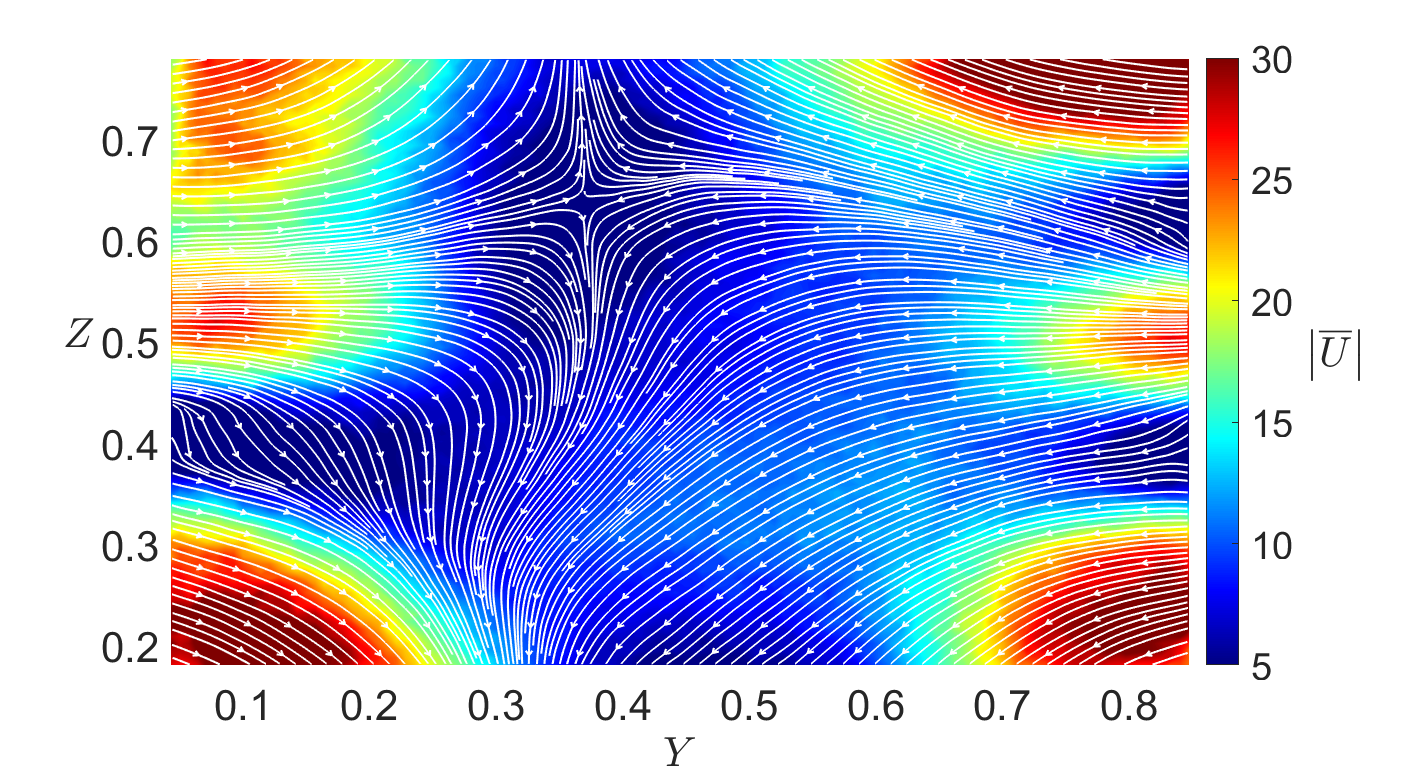}
\caption{\label{Fig2}
Spatial distributions of the mean velocity field $\meanU=[\meanU_y^2+\meanU_z^2]^{1/2}$ in the $YZ$ plane for turbulence generated by one oscillating grid (left panel) and two oscillating grids (right panel).
Streamlines (white) of the mean velocity field $\meanUU$ are superimposed on the distributions.
The velocity is measured in cm/s.
The coordinates $Y$ and $Z$ are normalized by $L_z=26$ cm.
}
\end{figure*}

\begin{figure*}[t!]
\centering
\includegraphics[width=6.0cm]{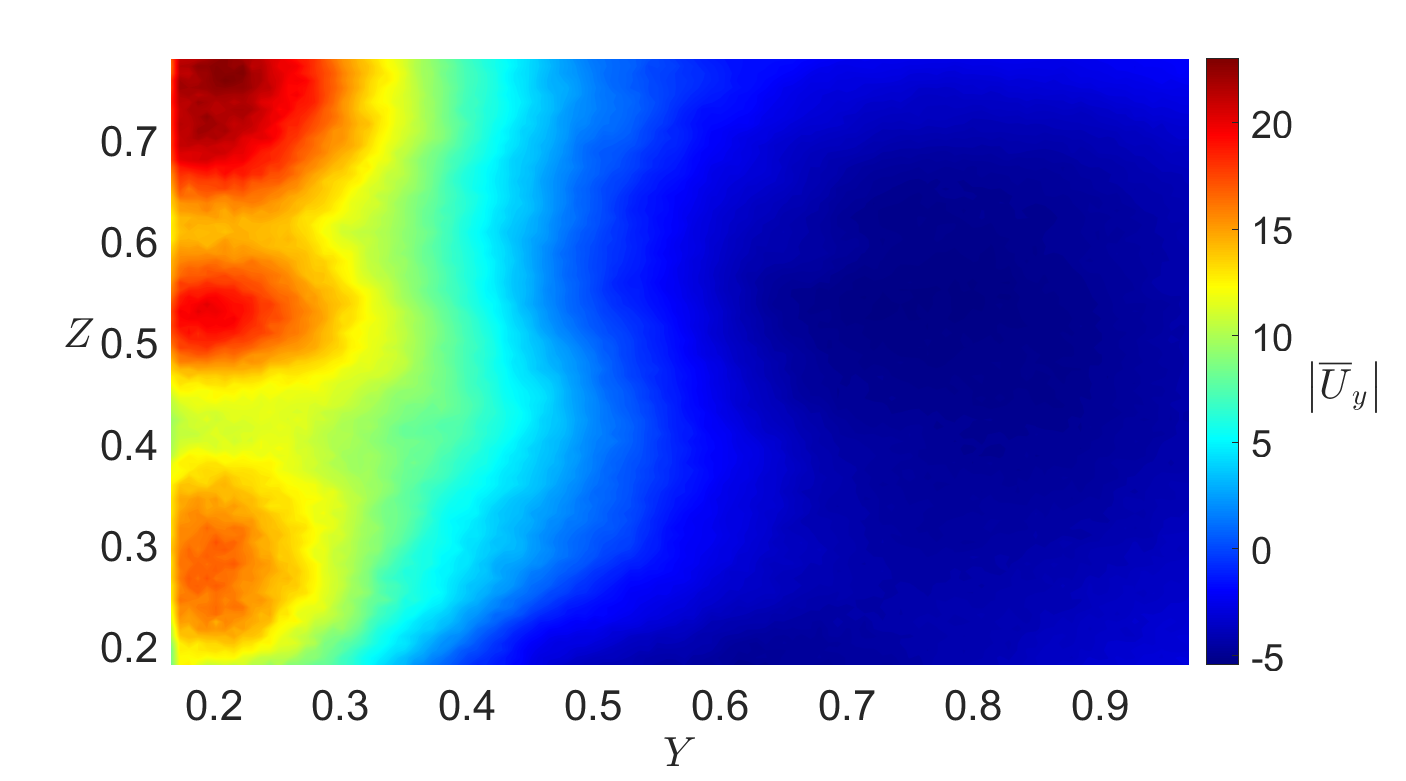}
\includegraphics[width=6.0cm]{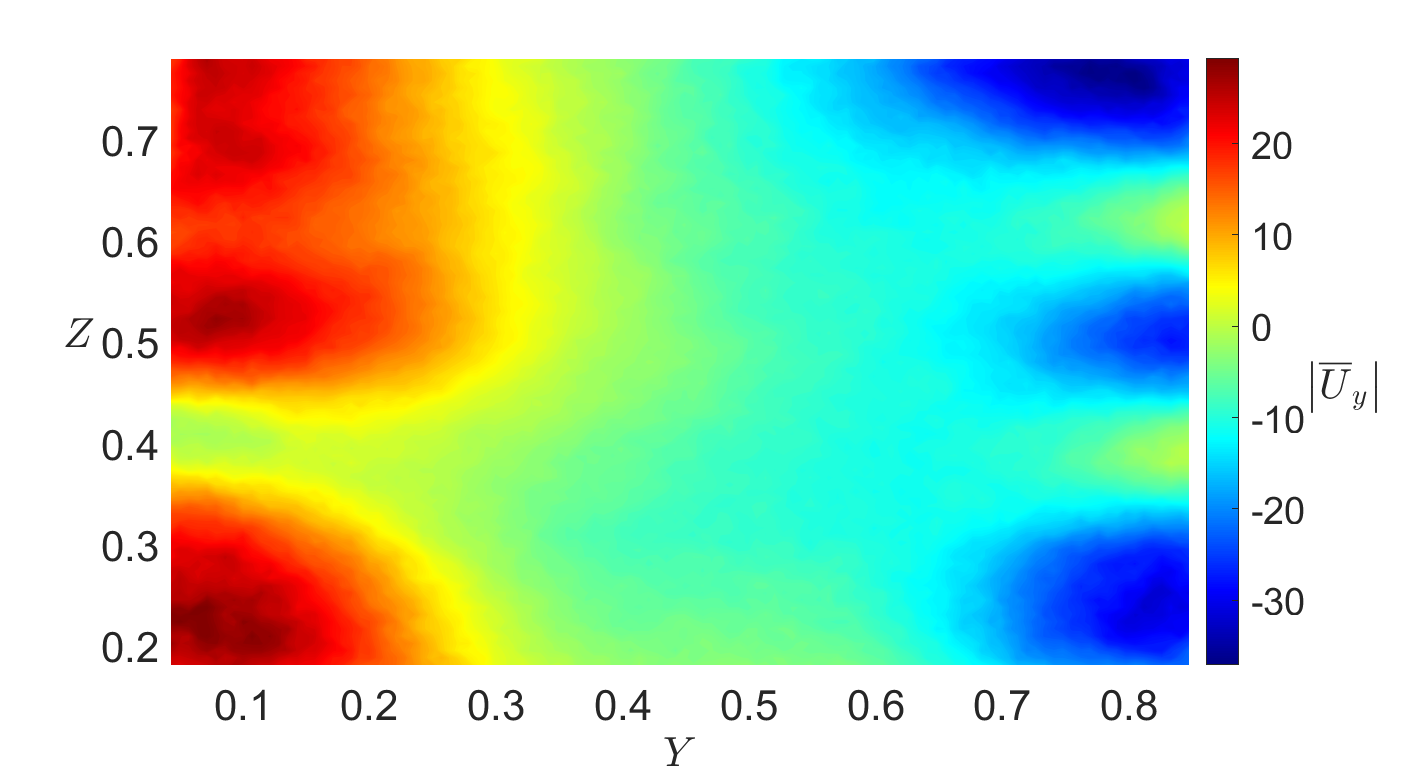}
\caption{\label{Fig2c}
Spatial distributions of the mean horizontal velocity field $\meanU_y$ in the $YZ$ plane for turbulence generated by one oscillating grid (left panel) and two oscillating grids (right panel).
The velocity is measured in cm/s.
The coordinates $Y$ and $Z$ are normalized by $L_z=26$ cm.
}
\end{figure*}

\begin{figure*}[t!]
\centering
\includegraphics[width=6.0cm]{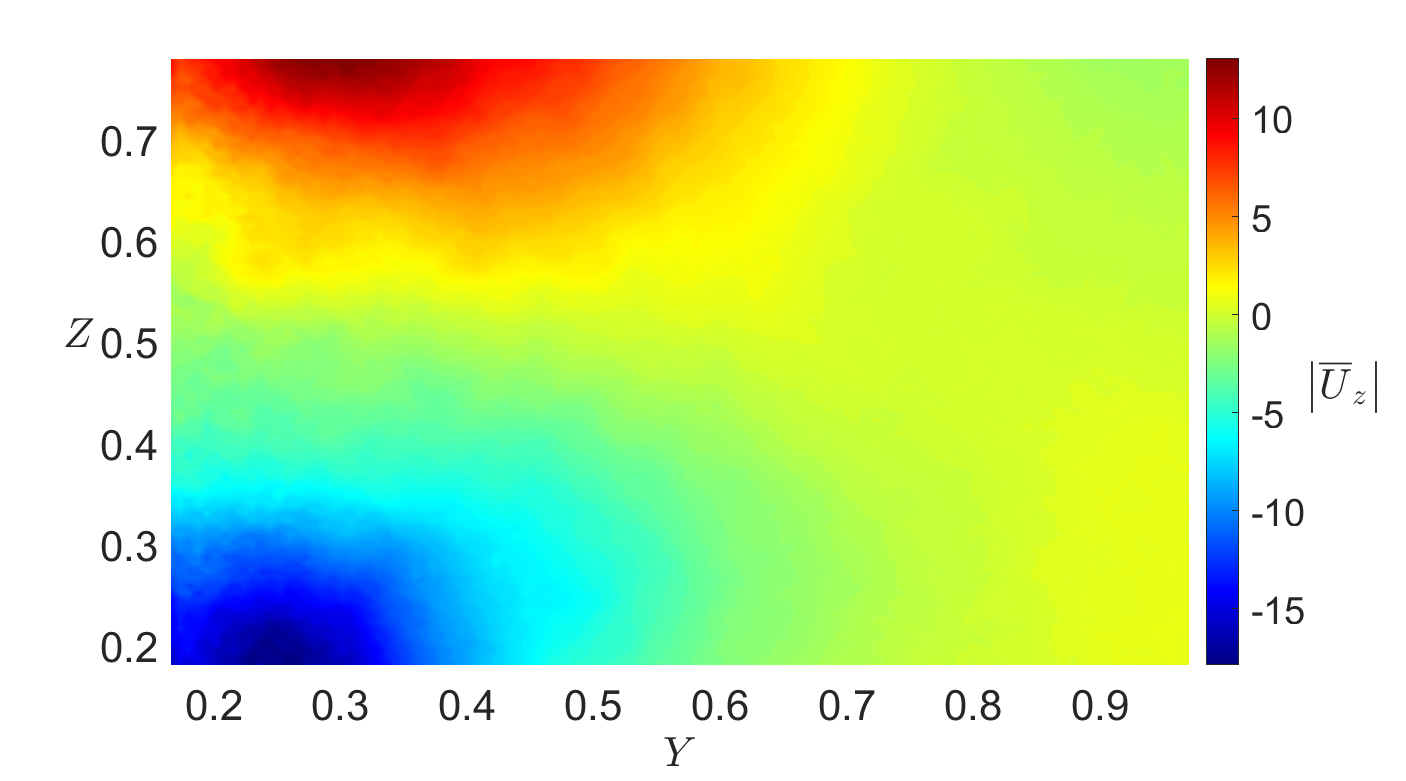}
\includegraphics[width=6.0cm]{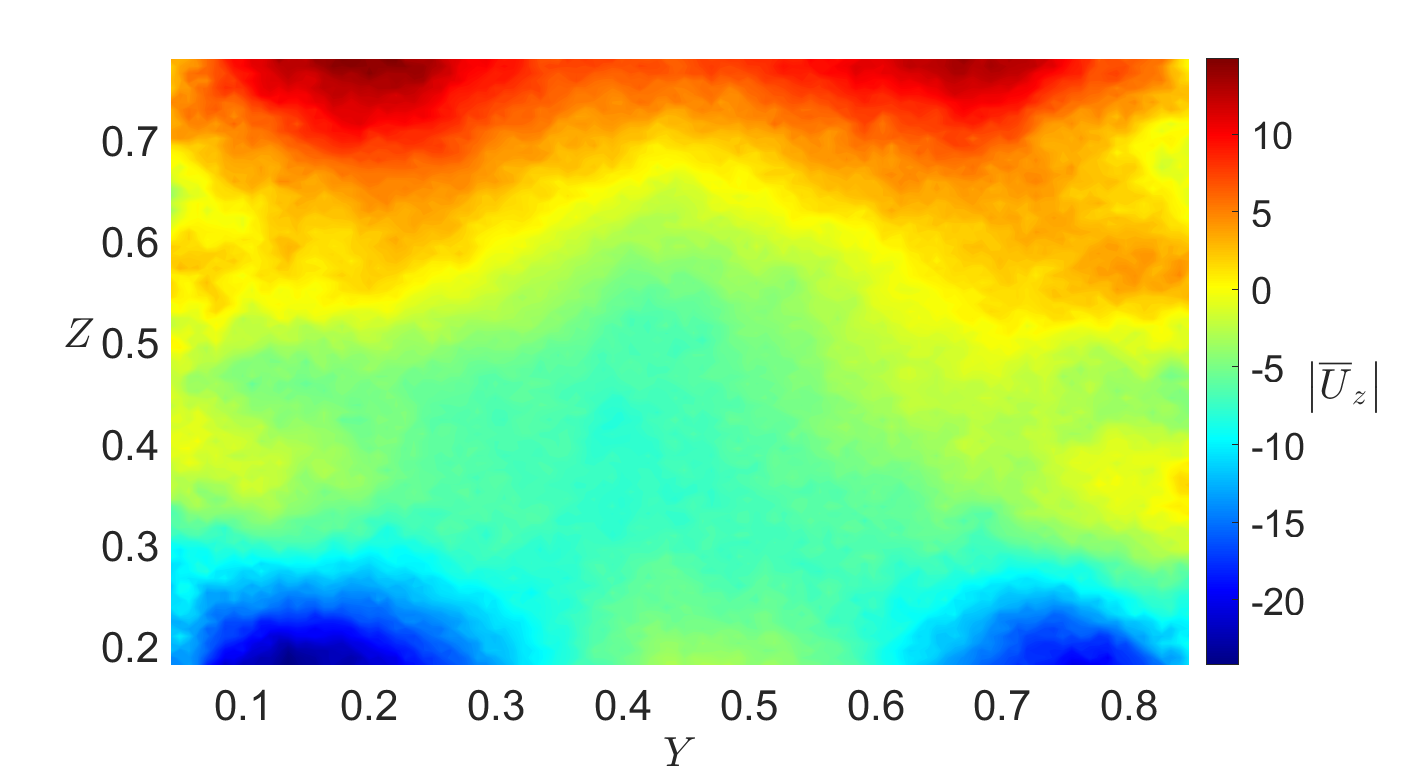}
\caption{\label{Fig2e}
Spatial distributions of the mean vertical velocity field $\meanU_z$ in the $YZ$ plane
for turbulence generated by one oscillating grid (left panel) and two oscillating grids (right panel).
The velocity is measured in cm/s.
The coordinates $Y$ and $Z$ are normalized by $L_z=26$ cm.
}
\end{figure*}

Figures~\ref{Fig2}--\ref{Fig2e} shows the spatial distributions of the mean velocity field $\meanUU=[\meanU_y^2+\meanU_z^2]^{1/2}$,
and its horizontal $\meanU_y$ and vertical $\meanU_z$ components, respectively, in the $YZ$ plane for inhomogeneous isothermal turbulence generated by one oscillating grid (left panel) and two oscillating grids (right panel). The mean velocity field produced by two oscillating grids is nearly symmetric with respect to the $Y$ axis.

\begin{figure*}[t!]
\centering
\includegraphics[width=6.0cm]{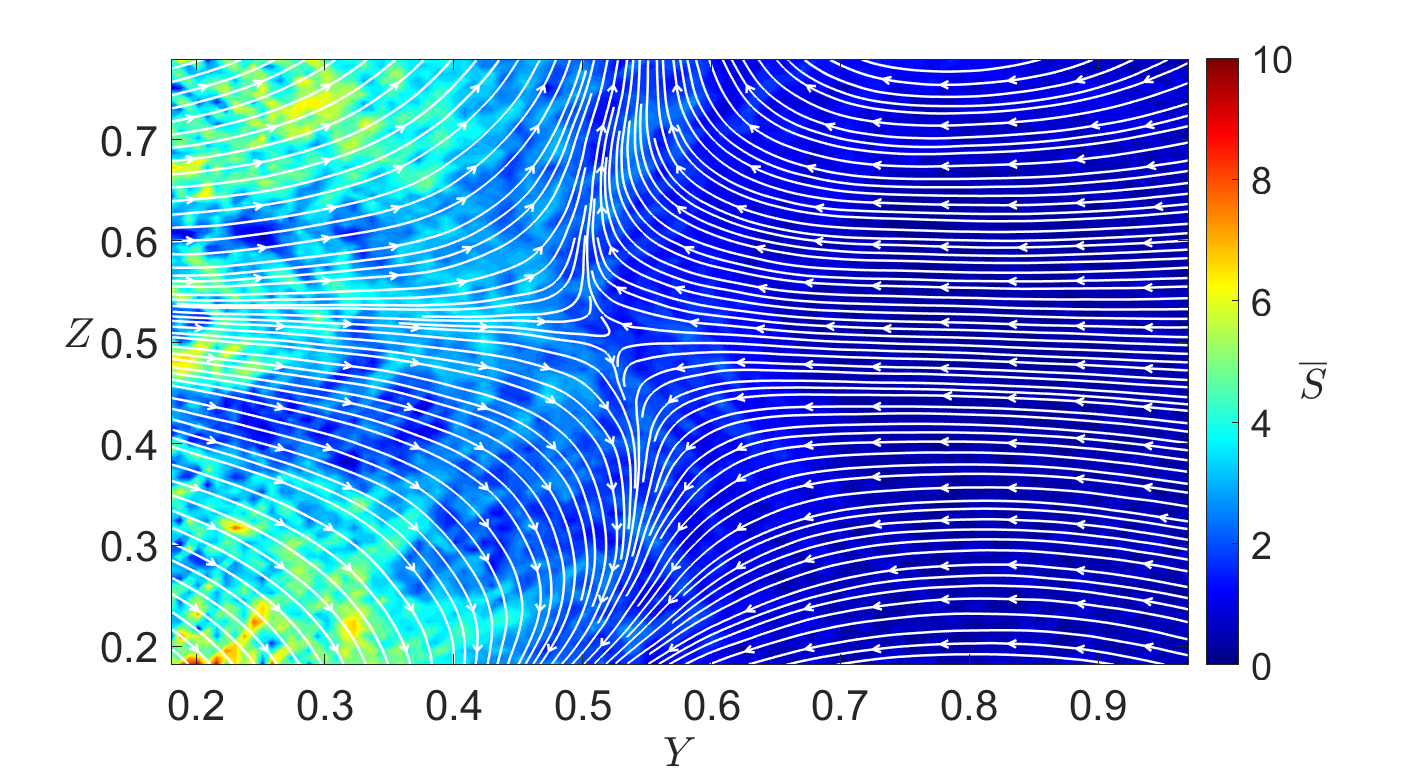}
\includegraphics[width=6.0cm]{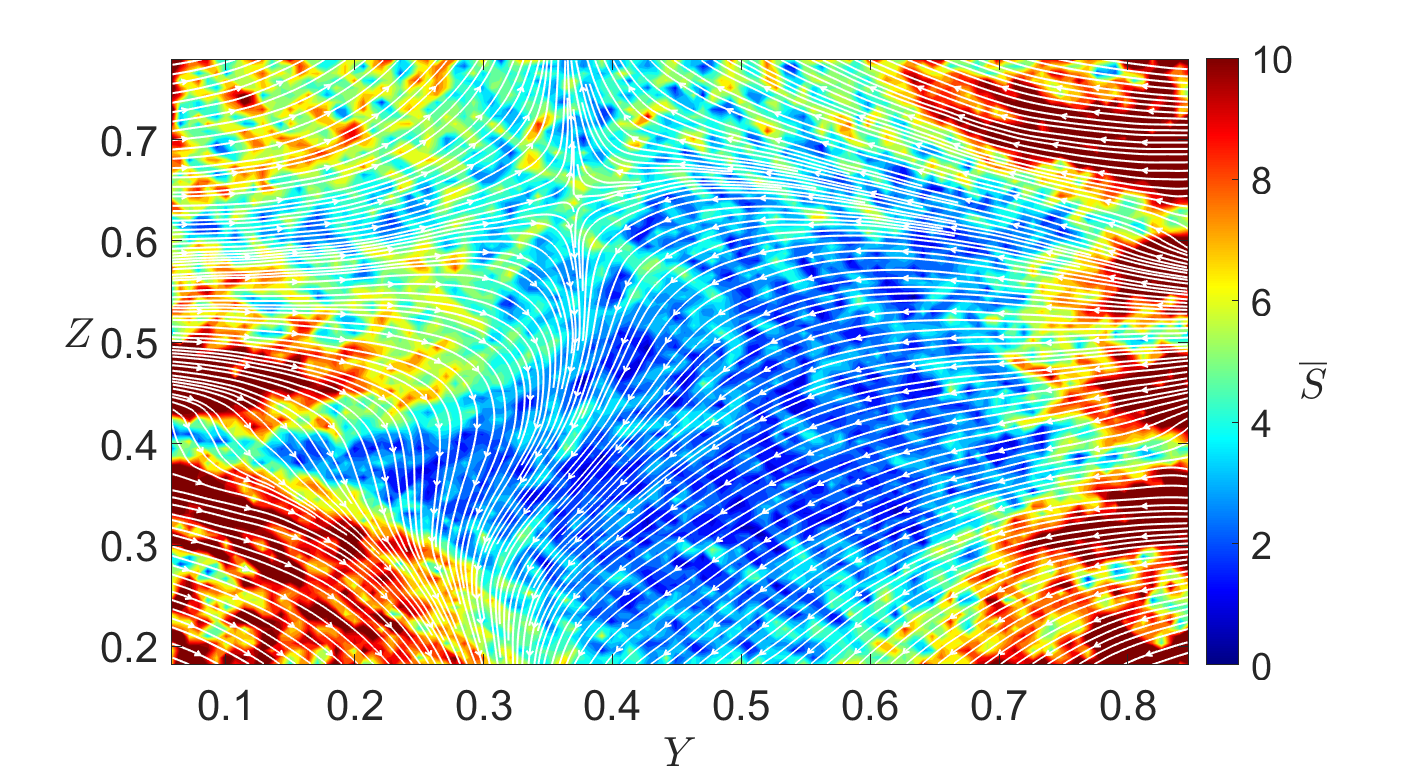}
\caption{\label{Fig3}
Spatial distributions of the mean velocity shear
$\meanS=\left[(\nabla_y \meanU_y)^2 + (\nabla_z \meanU_y)^2 + (\nabla_y \meanU_z)^2
+ (\nabla_z \meanU_z)^2\right]^{1/2}$
in the $YZ$ plane for turbulence generated by one oscillating grid (left panel) and two oscillating grids (right panel). Streamlines (white) of the mean velocity field $\meanUU$ are superimposed on the distributions. The coordinates $Y$ and $Z$ are normalized by $L_z=26$ cm, and the mean velocity shear $\meanS$ is measured in s$^{-1}$.
}
\end{figure*}

The large-scale velocity shear
$\meanS=\left[(\nabla_y \meanU_y)^2 + (\nabla_z \meanU_y)^2 + (\nabla_y \meanU_z)^2
+ (\nabla_z \meanU_z)^2\right]^{1/2}$
is stronger near the oscillating grids (see Fig.~\ref{Fig3}).
The maximum dimensionless shear parameter is approximately $\meanS^{\,({\rm max})} \tau_0 \approx 0.3$ for turbulence generated by one oscillating grid, where the characteristic turbulent time is $\tau_0 \approx 0.05$ s, and $\meanS^{\,({\rm max})} \tau_0 \sim 0.4$ for turbulence generated by two oscillating grids, where $\tau_0 \approx 0.033$ s.

\subsection{Characterisation of turbulence}
\label{subsect4.1}

In this Section, we shortly discuss the characterisation of turbulence in our experiments.
Dimensionless parameters characterising the turbulence and inertial particles for the
one-grid and two-grid systems are given in Table~\ref{table-2}.
Here $\ell_\nu= \ell_0$ Re$^{-3/4}$ is the Kolmogorov scale, d$_{\rm p} =10 \mu$m is the diameter of inertial particles,
$V_g$ is the terminal fall velocity of inertial particles, Re is the Reynolds number based on the integral scale $\ell_0$ of turbulence,
Re$_\lambda=\ell_\lambda u' /\nu$ is the Reynolds number based on the Taylor micro-scale $\ell_\lambda=[15 \nu \rho/\varepsilon_{_{K}}]^{1/2}$, where $\varepsilon_{_{K}}$ is the kinetic energy dissipation rate,  $\rho$ is the fluid density,
$u'=u_{\rm rms}/\sqrt{3}$,
St $= \tau_{\rm p}/\tau_0$ is the Stokes number based on the turbulent time $\tau_0$ at the integral scale,
St$_\nu= \tau_{\rm p}/\tau_\nu$ is the Stokes number based on the Kolmogorov time $\tau_\nu=\tau_0$ Re$^{-1/2}$.

\begin{table}
\caption{\label{table-2}
Dimensionless parameters characterising the turbulence and inertial particles for the
one-grid and two-grid systems}
\begin{tabular}{| c| c| c|}

\hline
Parameters                                                          &  one-grid system                        &  two-grid system \\
\hline
Integral-scale Reynolds number,   Re                  &   $120 - 500$                              &  $250- 750$   \\
\hline
Taylor-scale Reynolds number,   Re$_\lambda$   &   $14.3 - 29.1$                           &  $20.5   - 35.8$   \\
\hline
Integral-scale Stokes number,    St            &   $(5.3 - 8.2) \times 10^{-3}$           &  $(8.0 - 12.3) \times 10^{-3}$   \\
\hline
Kolmogorov-scale Stokes number,  St$_\nu$     & $(5.3 - 18.2) \times 10^{-2}$             & $(8.0 - 27.6) \times 10^{-2}$     \\
\hline
$\ell_\nu / d_{\rm p}$                         & $50 - 55$                                            & $17-24.4$     \\
\hline
$V_g / u_{\rm rms}  $                         & $(1.9 - 16.5) \times 10^{-2}$             & $(1.0 - 1.7) \times 10^{-2}$     \\
\hline
\end{tabular}
\end{table}

\begin{figure*}[t!]
\centering
\includegraphics[width=6.0cm]{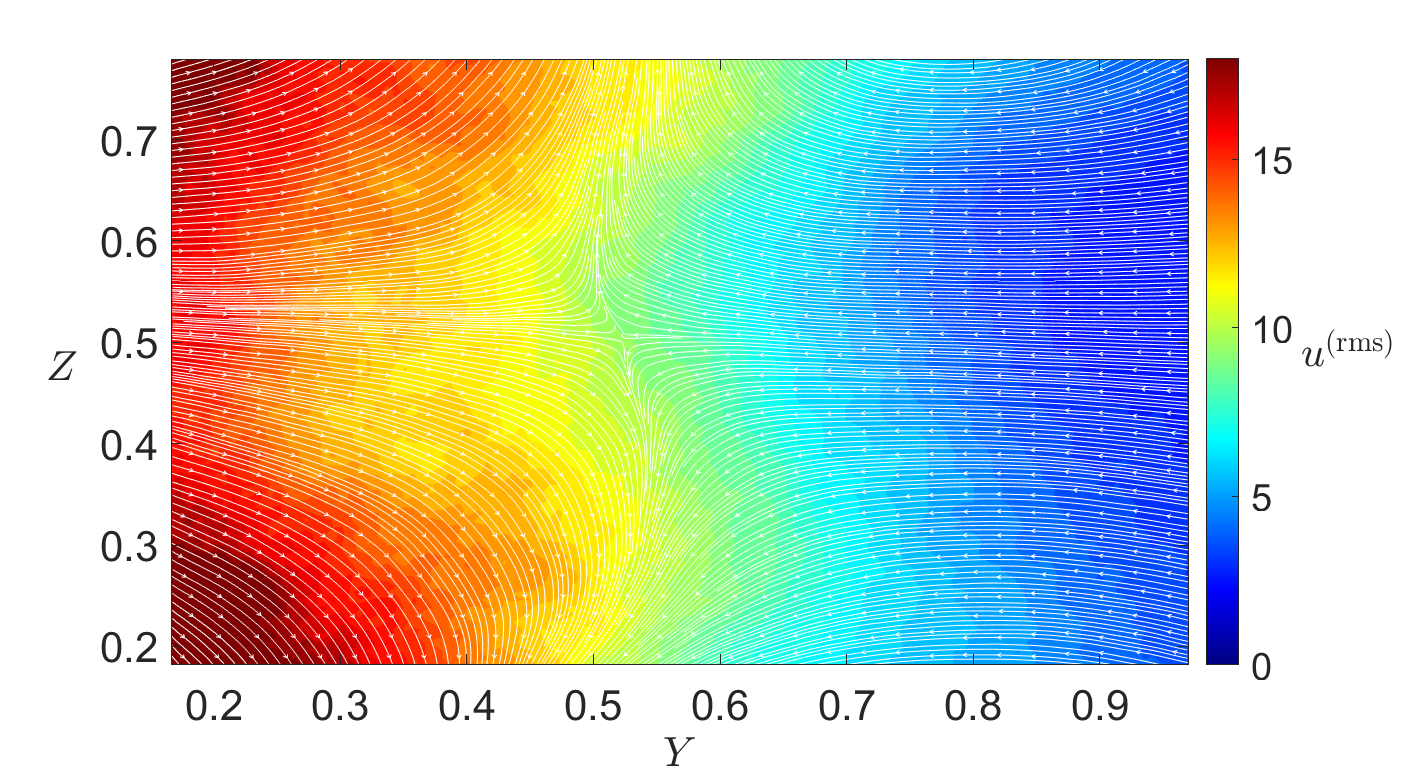}
\includegraphics[width=6.0cm]{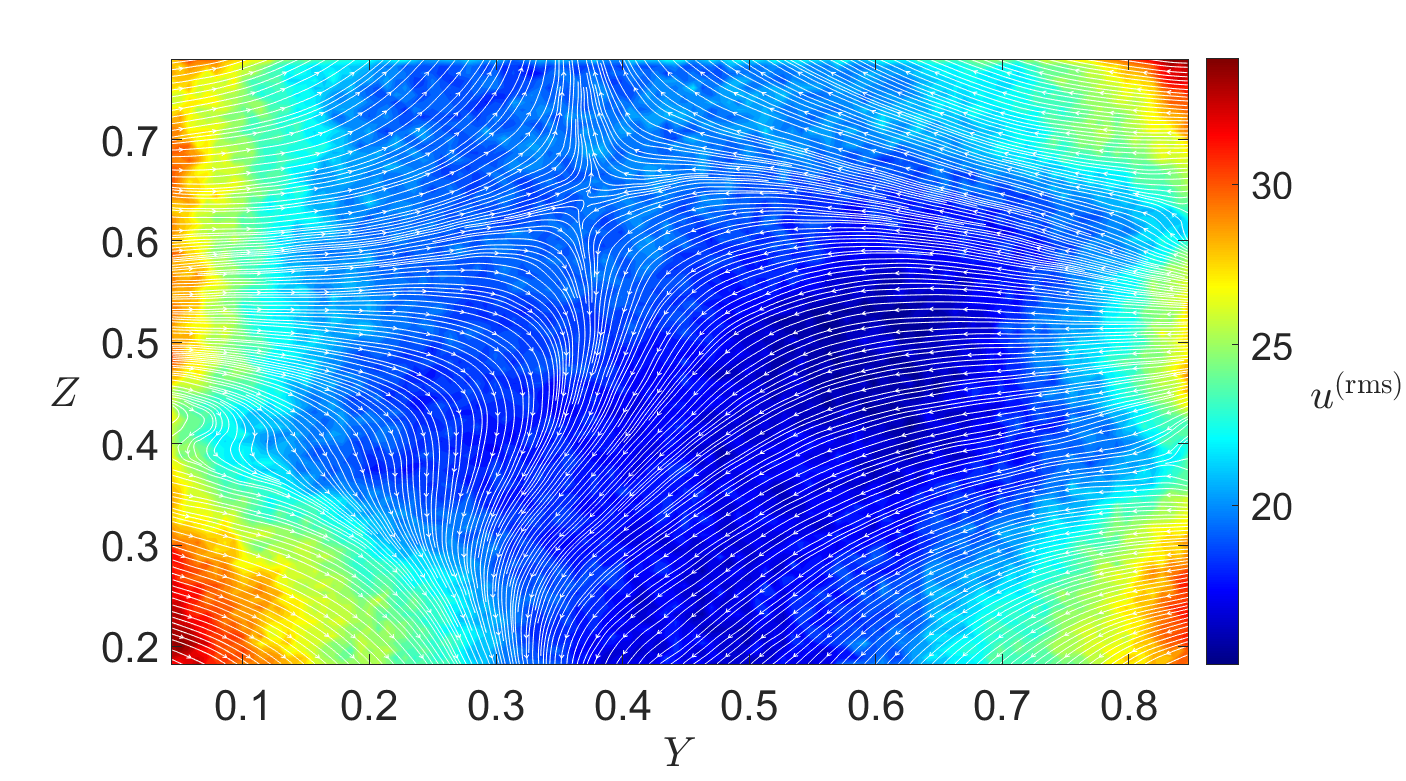}
\caption{\label{Fig4}
Spatial distributions of the turbulent velocity
$
u^{\rm (rms)} = \left[\langle u_y^2 \rangle + \langle u_z^2 \rangle\right]^{1/2}
$
in the $YZ$ plane for turbulence generated by one oscillating grid (left panel) and two oscillating grids (right panel). Streamlines (white) of the mean velocity field $\meanUU$ are superimposed on the distributions. The turbulent velocity is measured in cm/s, and the coordinates $Y$ and $Z$ are normalized by $L_z=26$ cm.
}
\end{figure*}

\begin{figure*}[t!]
\centering
\includegraphics[width=6.0cm]{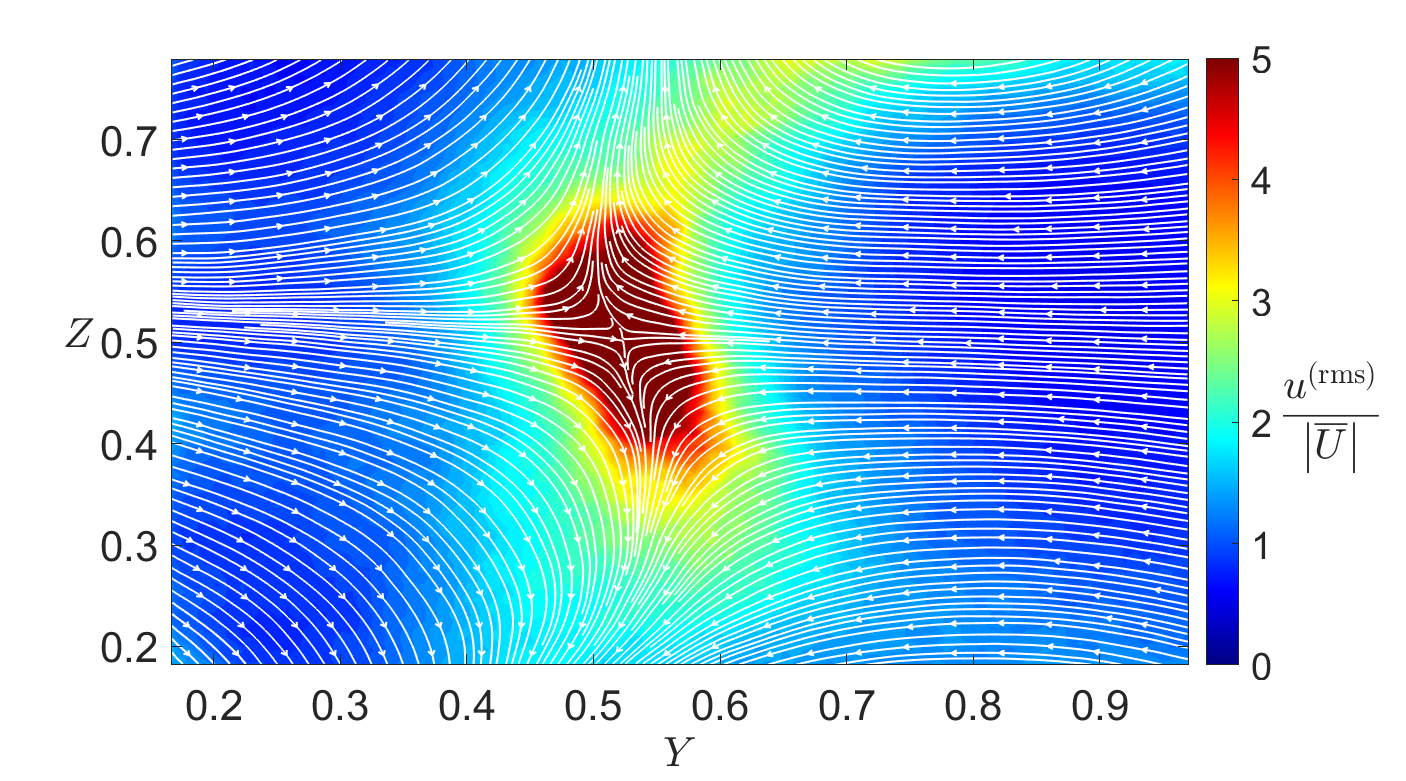}
\includegraphics[width=6.0cm]{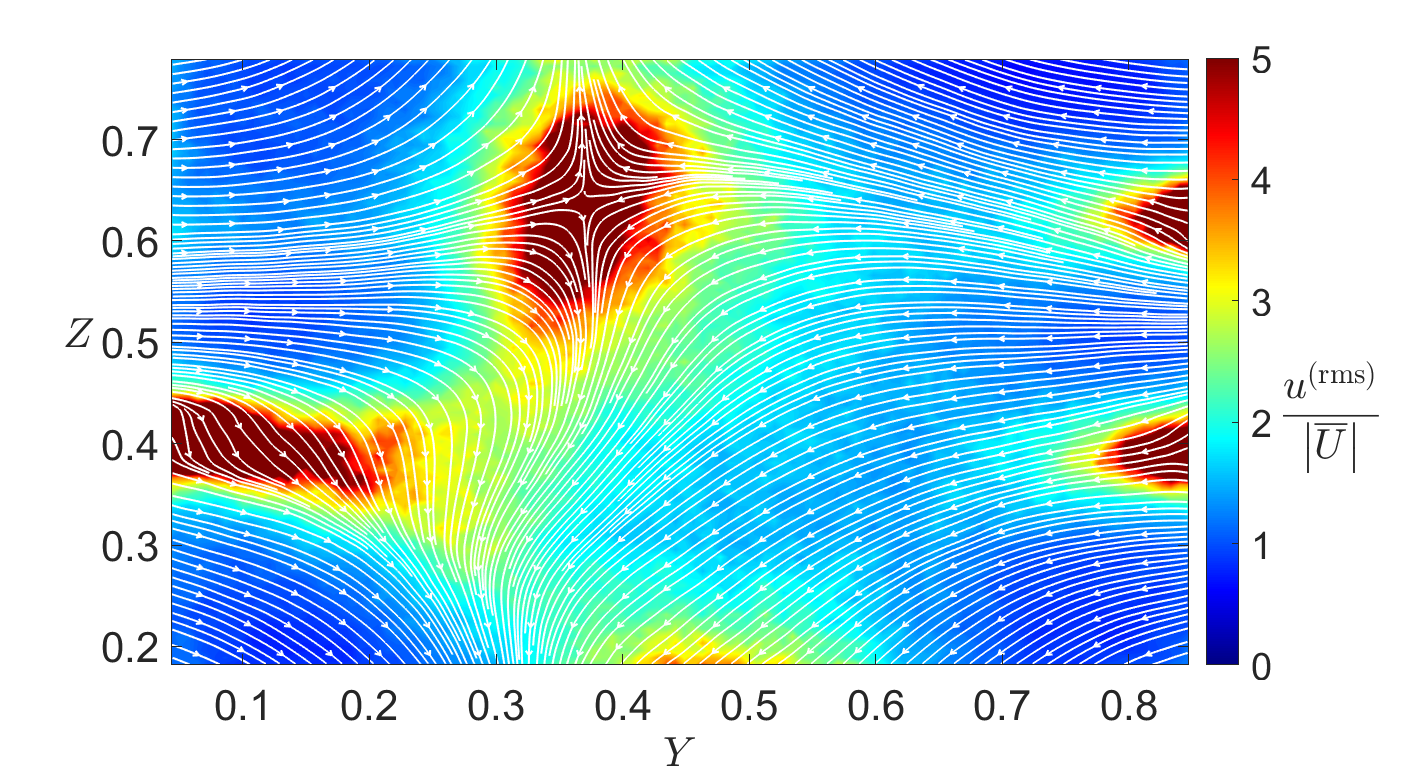}
\caption{\label{Fig5}
Spatial distributions of the ratio $u^{\rm (rms)} / |\meanUU|$
in the $YZ$ plane for turbulence generated by one oscillating grid (left panel) and two oscillating grids (right panel). Streamlines (white) of the mean velocity field $\meanUU$ are superimposed on the distributions. The coordinates $Y$ and $Z$ are normalized by $L_z=26$ cm.
}
\end{figure*}

\begin{figure*}[t!]
\centering
\includegraphics[width=6.0cm]{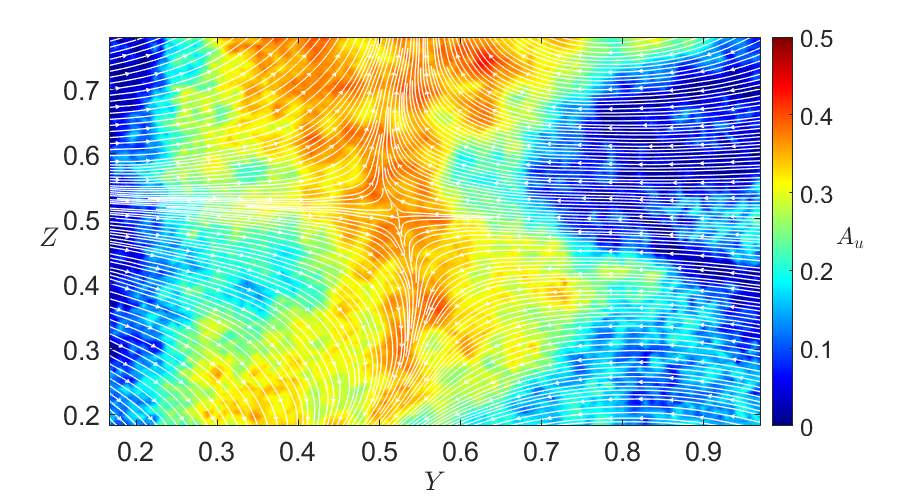}
\includegraphics[width=6.0cm]{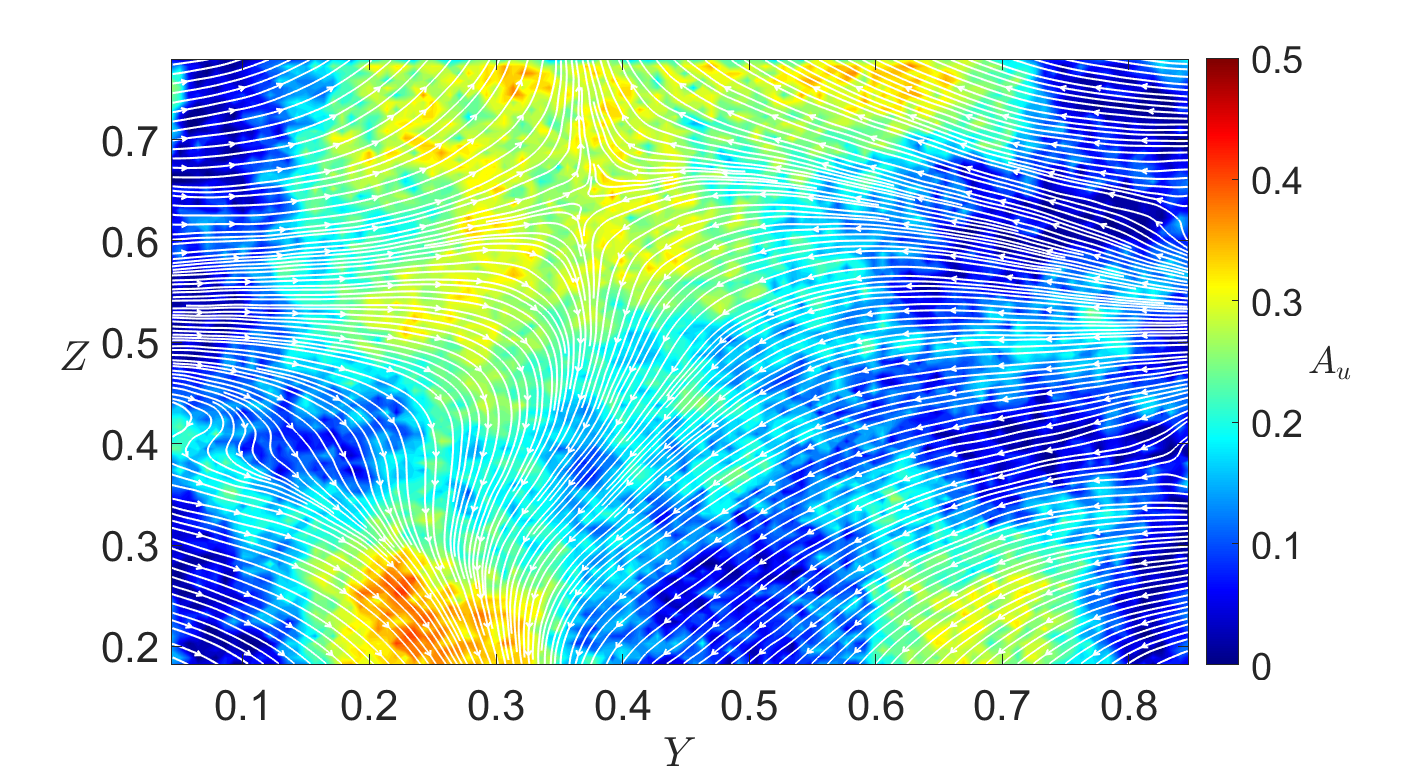}
\caption{\label{Fig6}
Spatial distributions of the anisotropy of turbulent velocity field $A_u=|u_z^{\rm (rms)} / u_y^{\rm (rms)} -1|$
in the $YZ$ plane for a turbulence produced by one oscillating grid  (left panel) and by two oscillating grids (right panel).
The streamlines (white) of the mean velocity $\meanUU$ are also superimposed on this distribution.
The coordinates $Y$ and $Z$ are normalized by $L_z=26$ cm.
}
\end{figure*}

\begin{figure*}[t!]
\centering
\includegraphics[width=6.0cm]{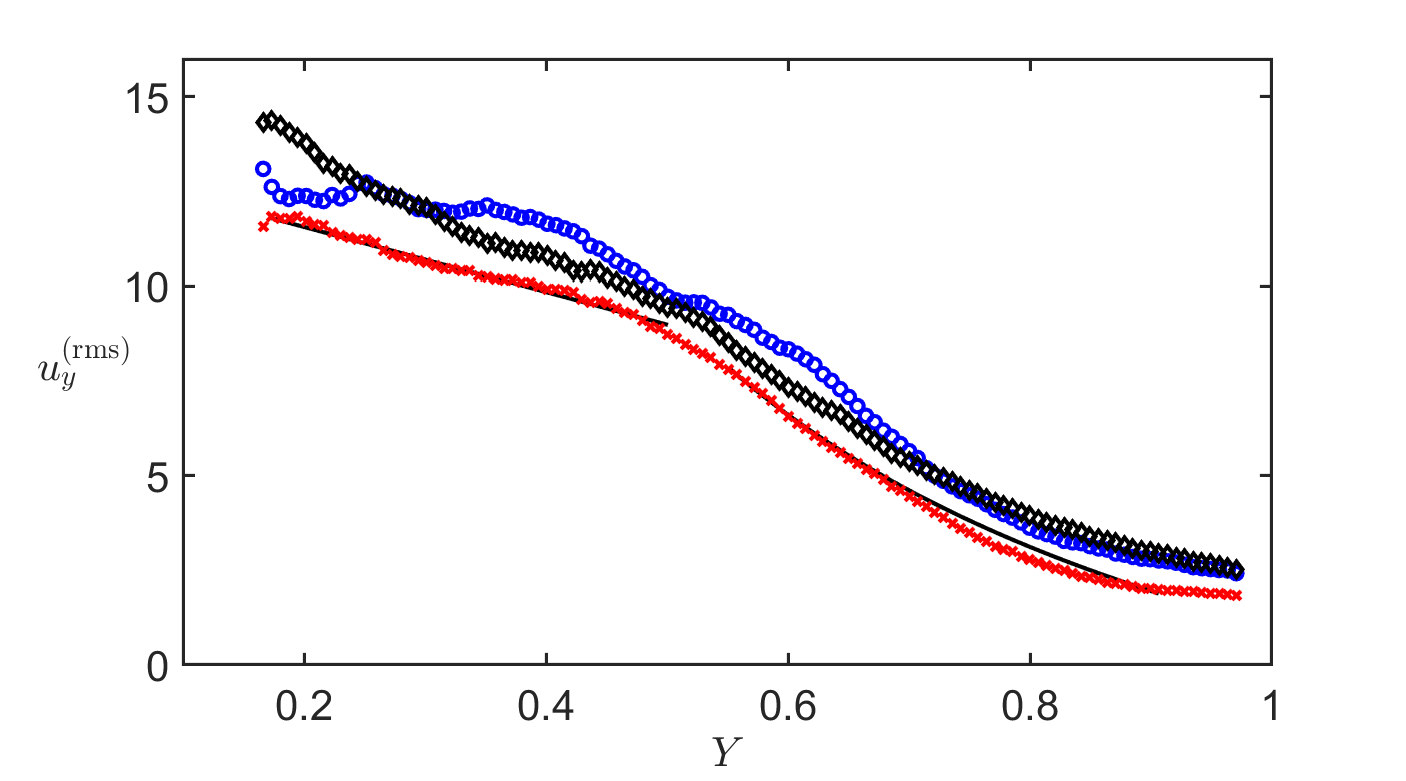}
\includegraphics[width=6.0cm]{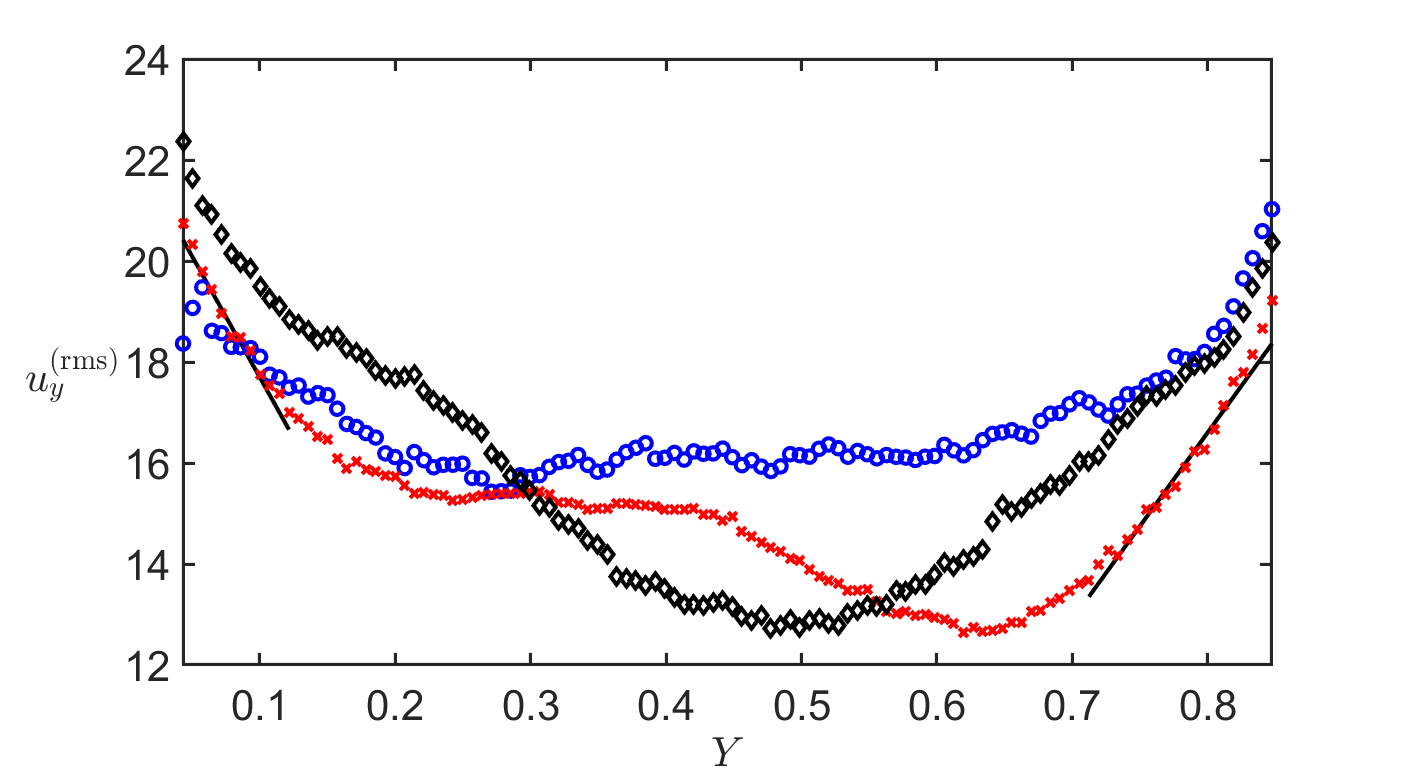}
\caption{\label{Fig7}
The horizontal turbulent velocity $u_y^{\rm (rms)}(Y)$ versus  the horizontal coordinate $Y$
in the core flow averaged over different ranges of $Z$
for turbulence generated by one oscillating grid (left panel) and two oscillating grids (right panel).
Solid lines show the fitting curves.
In the left panel, the averaging is performed over $Z =$ 5.2--9.3 cm (blue circles), 9.2--14.3 cm (red crosses), and 14.3--18.6 cm (black diamonds). In the right panel, the averaging is performed over $Z =$ 5.3--9.4 cm (blue circles), 9.4--14.2 cm (red crosses), and 14.2--18.4 cm (black diamonds). The velocity is measured in cm/s, and the coordinate $Y$ is normalized by $L_z=26$ cm.
}
\end{figure*}

\begin{figure*}[t!]
\centering
\includegraphics[width=6.0cm]{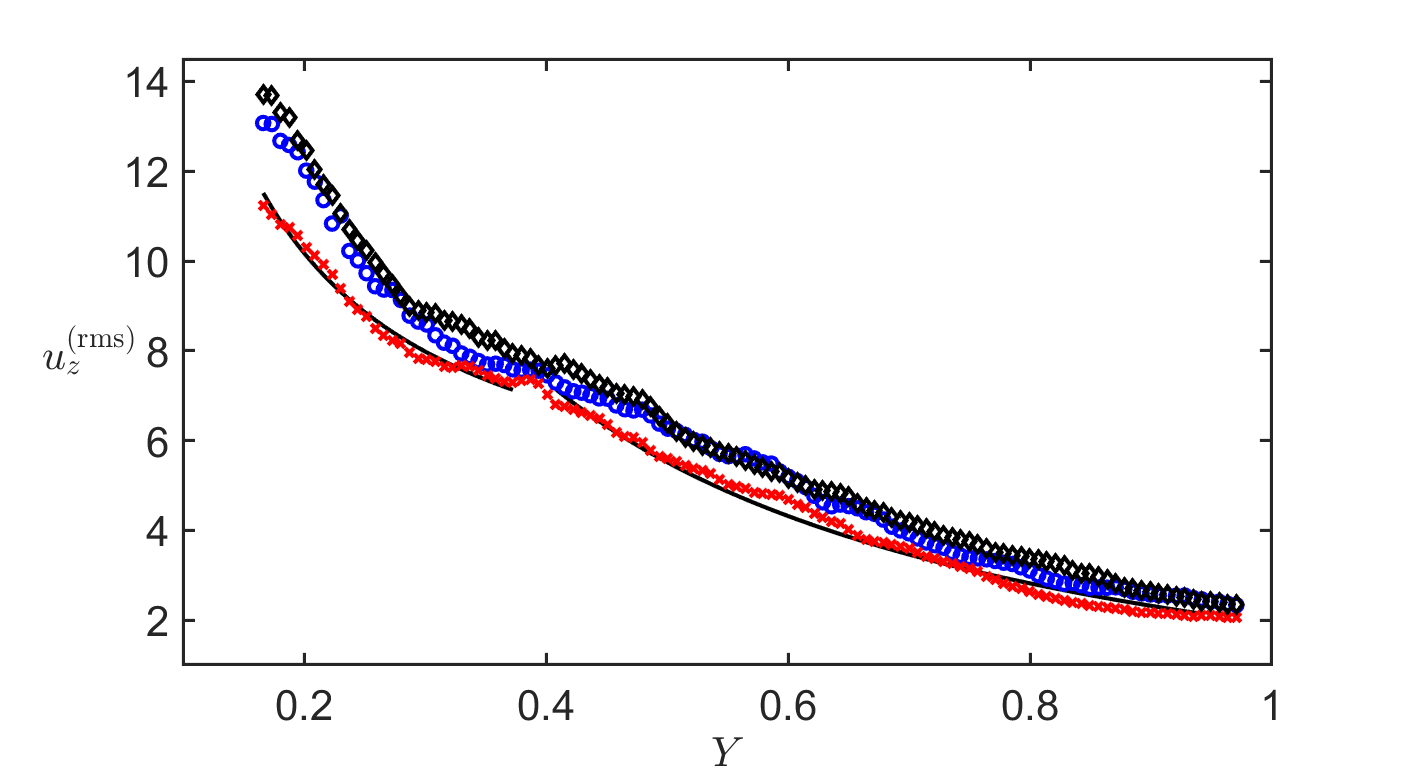}
\includegraphics[width=6.0cm]{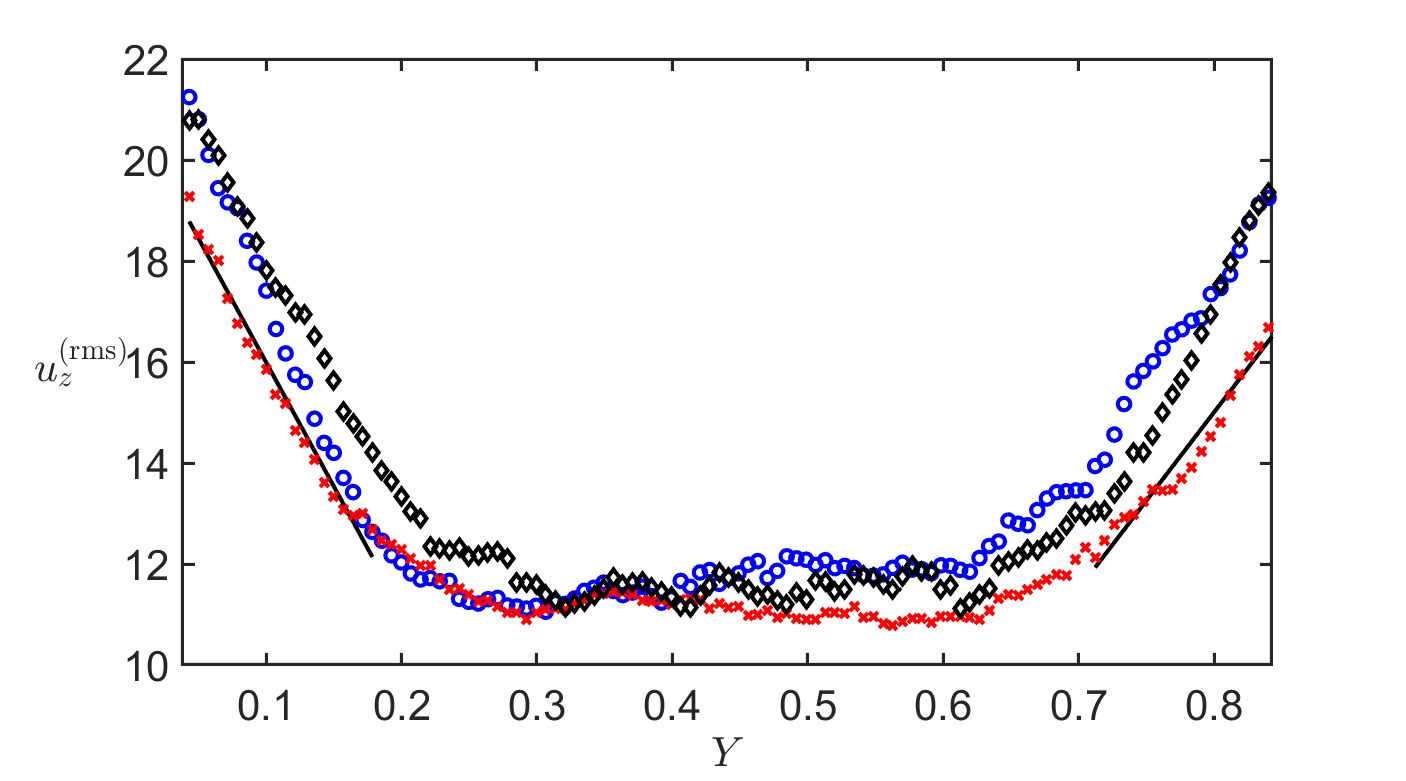}
\caption{\label{Fig8}
The vertical turbulent velocity
$u_z^{\rm (rms)}(Y)$
versus the horizontal coordinate $Y$ in the core flow averaged over different ranges of $Z$ for turbulence generated by one oscillating grid (left panel) and two oscillating grids (right panel). Solid lines show the fitting curves. In the left panel, the averaging is performed over $Z =$ 5.2--9.3 cm (blue circles), 9.2--14.3 cm (red crosses), and 14.3--18.6 cm (black diamonds). In the right panel, the averaging is performed over $Z =$ 5.3--9.4 cm (blue circles), 9.4--14.2 cm (red crosses), and 14.2--18.4 cm (black diamonds). The velocity is measured in cm/s, and the coordinate $Y$ is normalized by $L_z=26$ cm.
}
\end{figure*}

\begin{figure*}[t!]
\centering
\includegraphics[width=6.0cm]{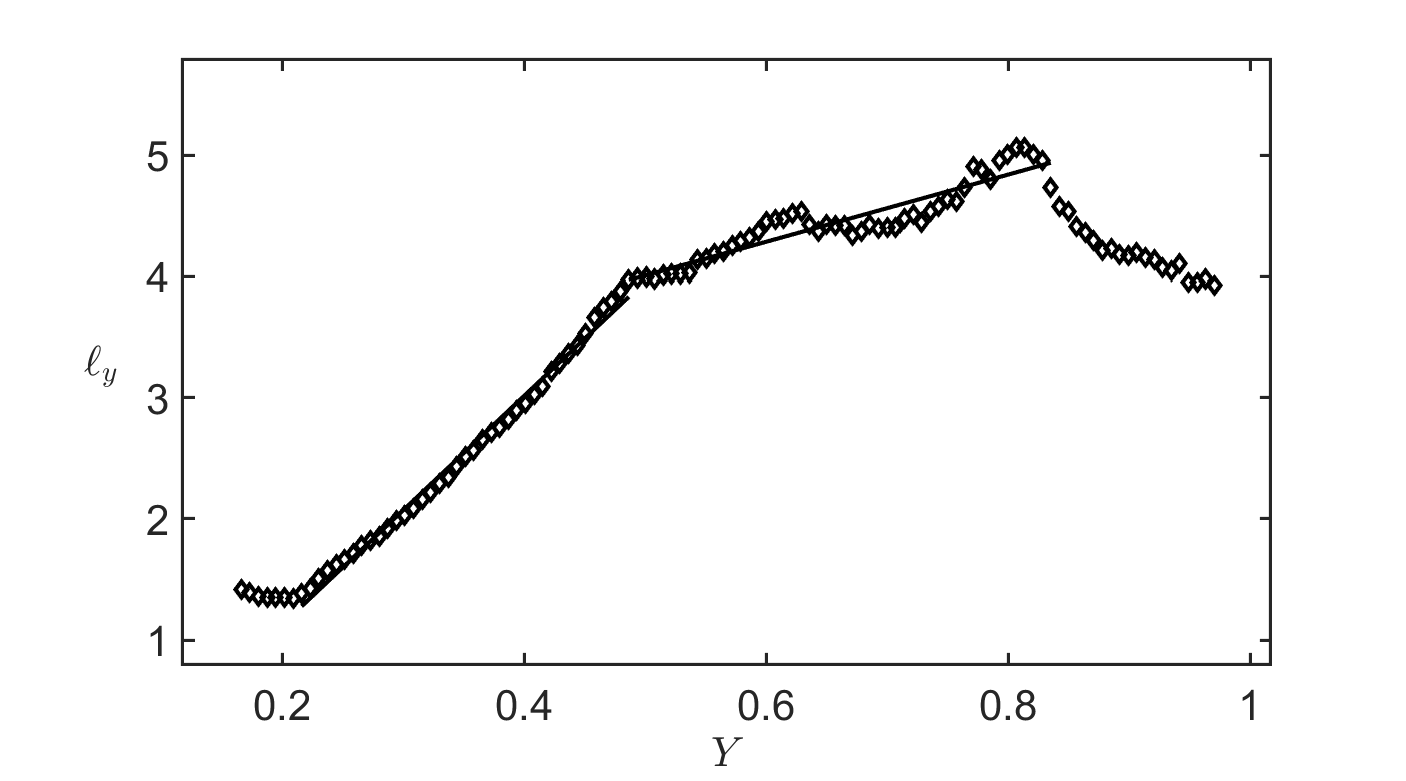}
\includegraphics[width=6.0cm]{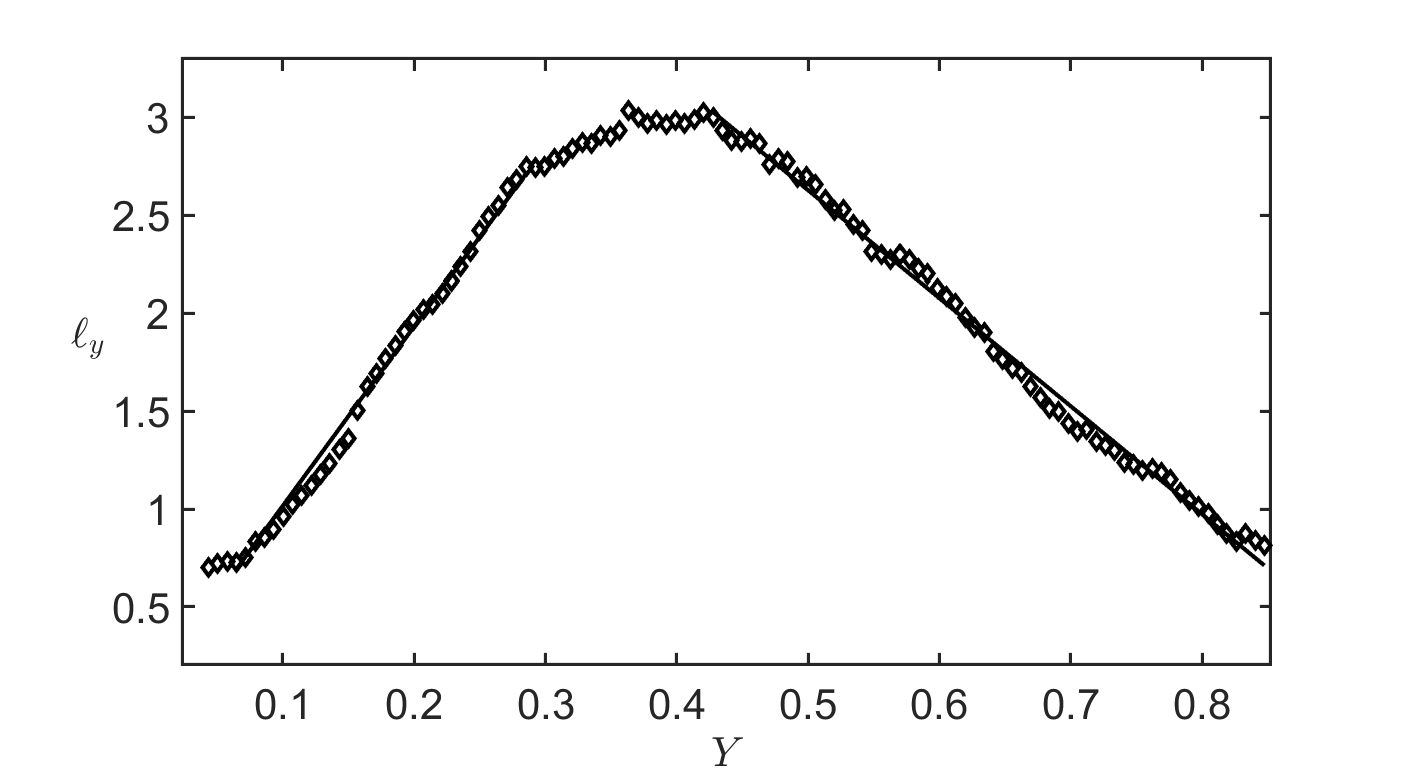}
\caption{\label{Fig9}
The horizontal integral turbulence scale
$\ell_y(Y)$
versus the horizontal coordinate $Y$ in the core flow averaged over $Z$ for turbulence generated by one oscillating grid (left panel) and two oscillating grids (right panel). Solid lines show the fitting curves. The integral turbulence scale is measured in cm, and the coordinate $Y$ is normalized by $L_z=26$ cm.
}
\end{figure*}

\begin{figure*}[t!]
\centering
\includegraphics[width=6.0cm]{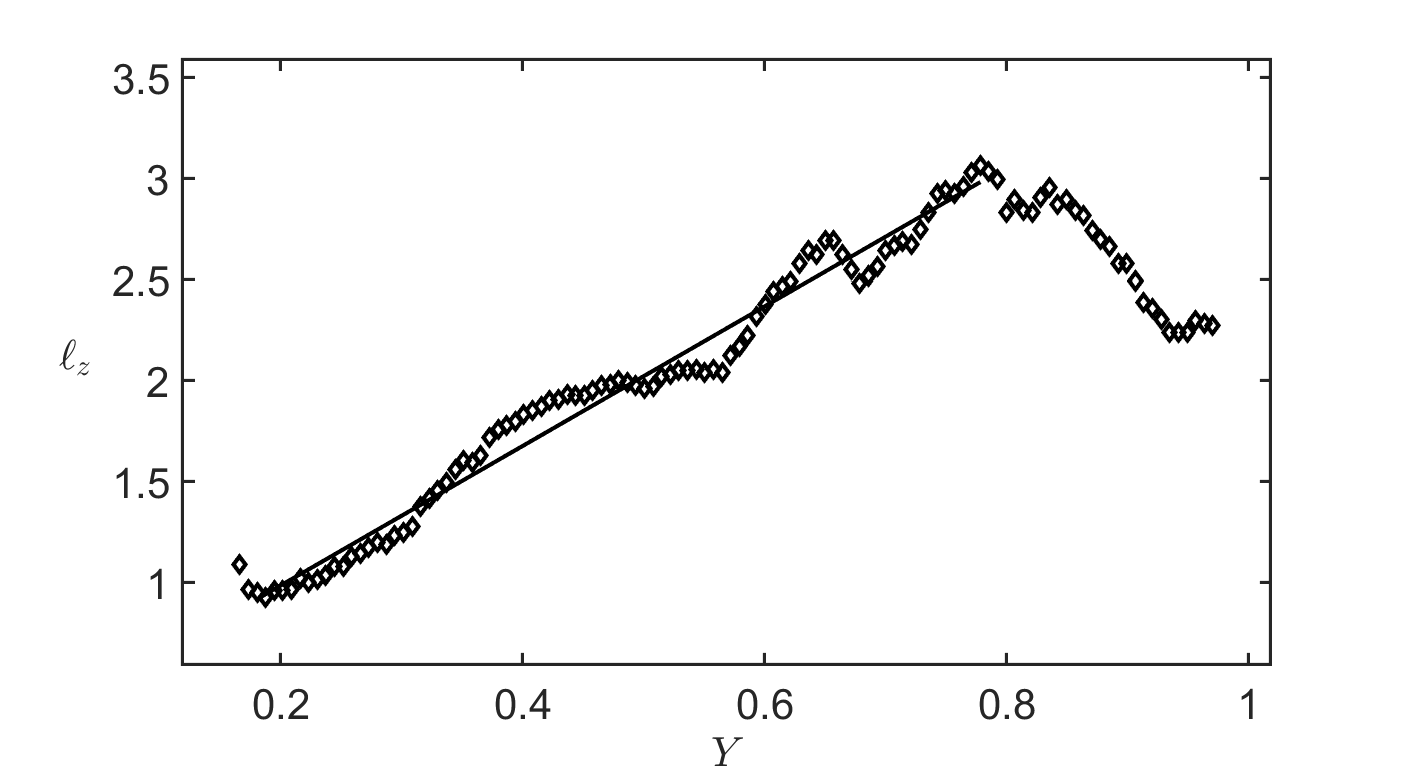}
\includegraphics[width=6.0cm]{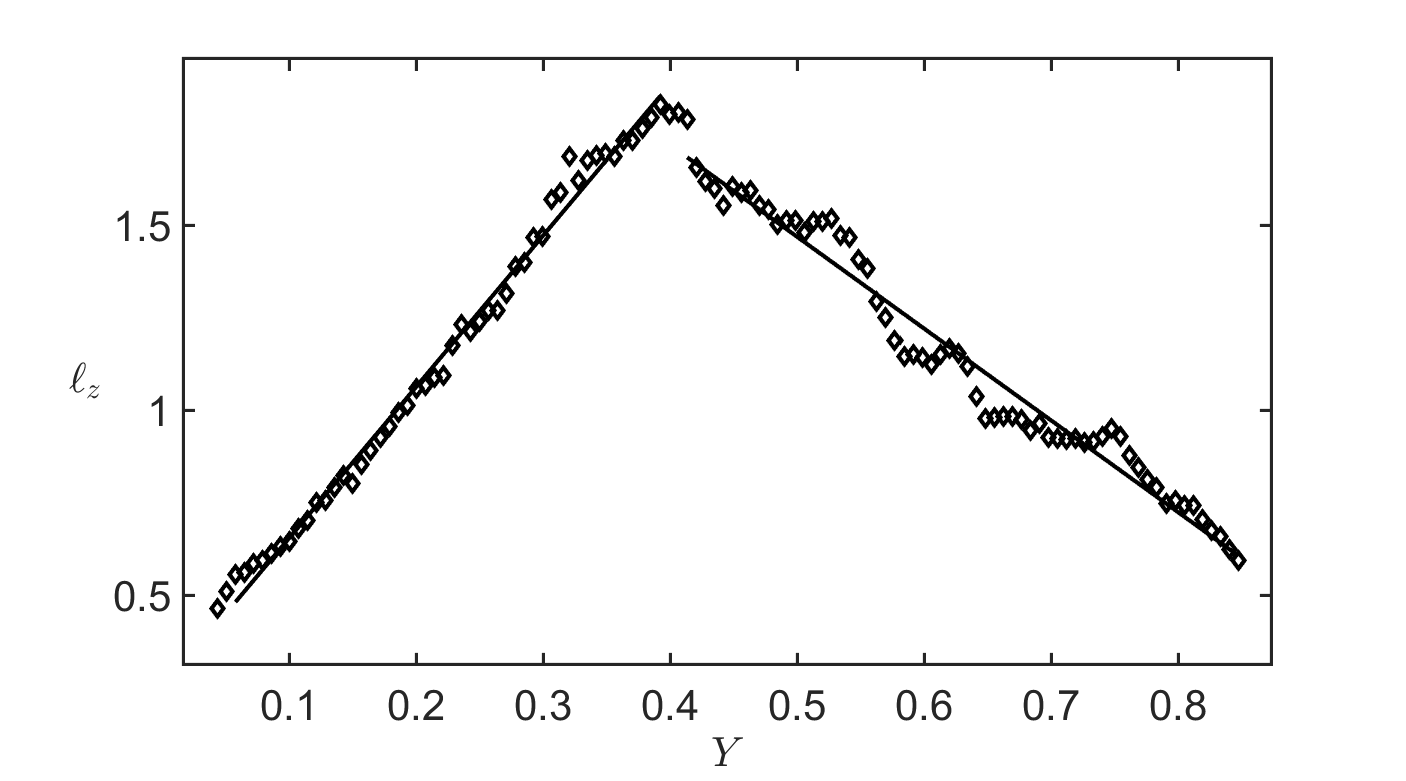}
\caption{\label{Fig10}
The vertical integral turbulence scale
$\ell_z(Y)$
versus the horizontal coordinate $Y$ in the core flow averaged over $Z$ for turbulence generated by one oscillating grid (left panel) and two oscillating grids (right panel). Solid lines show the fitting curves. The integral turbulence scale is measured in cm, and the coordinate $Y$ is normalized by $L_z=26$ cm.
}
\end{figure*}

\begin{figure*}[t!]
\centering
\includegraphics[width=6.0cm]{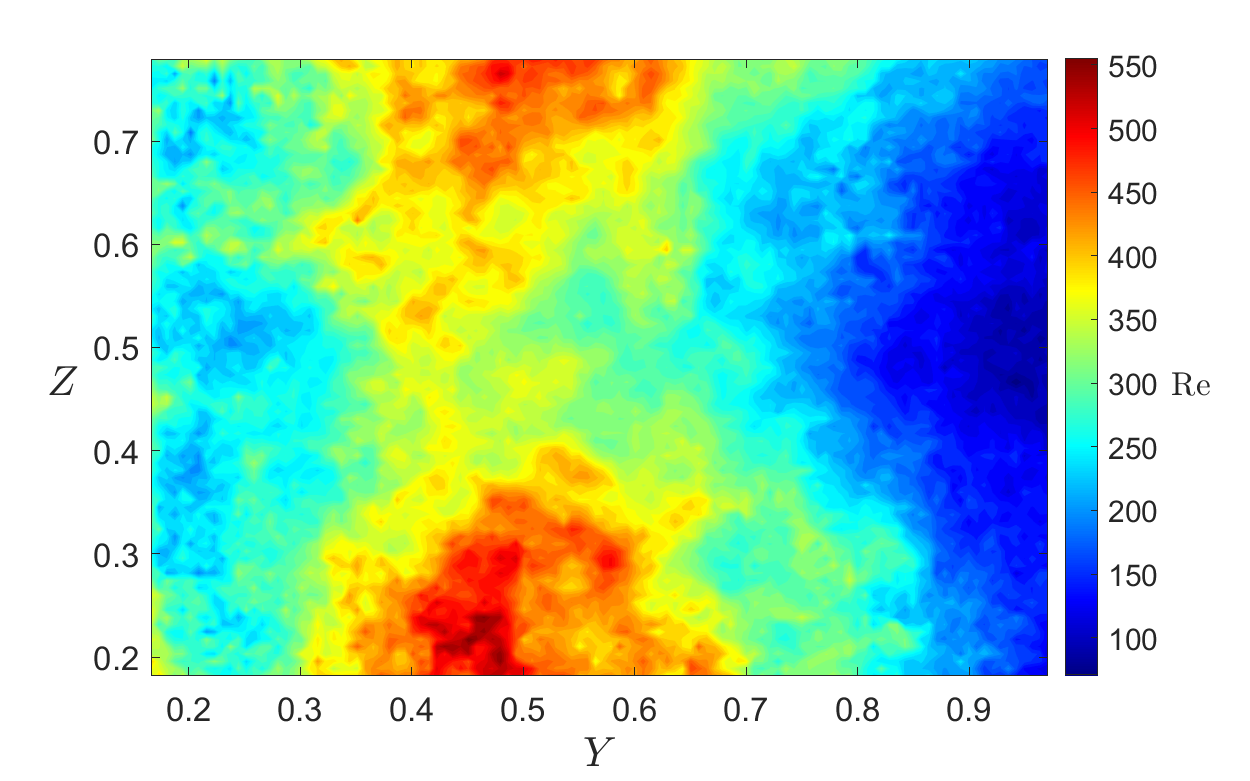}
\includegraphics[width=6.0cm]{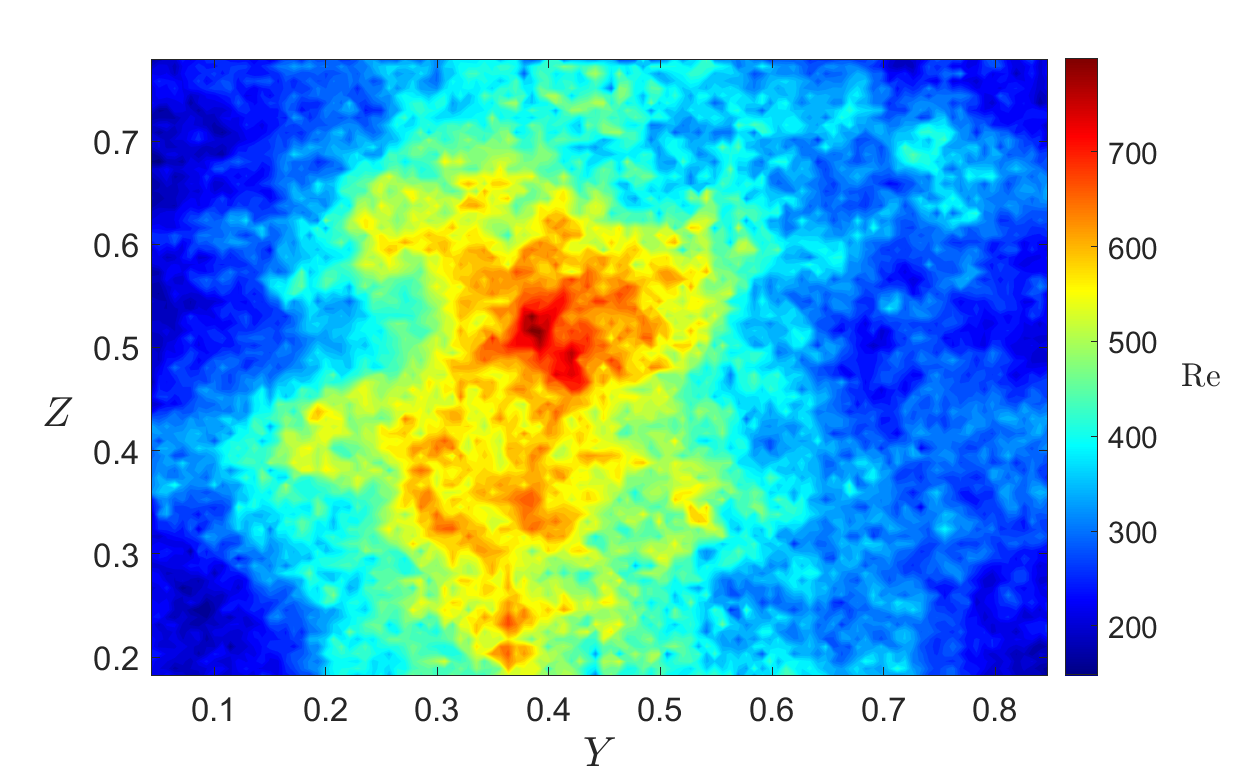}
\caption{\label{Fig11}
Spatial distributions of the Reynolds number
${\rm Re}=[u^{\rm (rms)}_y \, \ell_y + 2 u^{\rm (rms)}_z \, \ell_z]/\nu$
in the $YZ$ plane for turbulence generated by one oscillating grid (left panel) and two oscillating grids (right panel). The coordinates $Y$ and $Z$ are normalized by $L_z=26$ cm.
}
\end{figure*}

Figures~\ref{Fig4}--\ref{Fig11} present the basic turbulent characteristics in the $YZ$ plane:
\begin{itemize}
\item the turbulent velocity
$u^{\rm (rms)} = \left[\langle u_y^2 \rangle + \langle u_z^2 \rangle\right]^{1/2}$
(see Fig.~\ref{Fig4});
\item the ratio between turbulent and mean velocities
$u^{\rm (rms)} / |\meanUU|$
(see Fig.~\ref{Fig5});
\item the turbulence anisotropy parameter
$A_u=|u_z^{\rm (rms)} / u_y^{\rm (rms)} -1|$
(see Fig.~\ref{Fig6});
\item the horizontal profiles of the turbulent velocity components
$u_y^{\rm (rms)}(Y)$
and
$u_z^{\rm (rms)}(Y)$
(see Figs.~\ref{Fig7}--\ref{Fig8}), together with the integral turbulence length scales
$\ell_y(Y)$ and $\ell_z(Y)$
(see Figs.~\ref{Fig9}--\ref{Fig10});
\item the Reynolds number based on the integral turbulence length scale (see Fig.~\ref{Fig11})
\begin{eqnarray}
{\rm Re}={1 \over \nu} \, [u^{\rm (rms)}_y \, \ell_y + 2 u^{\rm (rms)}_z \, \ell_z] .
\label{RE1}
\end{eqnarray}
\end{itemize}

Generally, the Reynolds number  based on the integral turbulence length scale
is estimated as ${\rm Re}=\tau_0 \langle {\bm u}^2 \rangle/\nu$, where
$\langle {\bm u}^2 \rangle = \langle u_x^2 \rangle
+ \langle u_y^2 \rangle + \langle u_z^2 \rangle$ and $\tau_0$ is the characteristic turbulent time in the integral scale.
The estimate~(\ref{RE1}) assumes similar turbulent time scales in the $X$, $Y$, and $Z$ directions,
$\tau_0 = \ell_x/u_x^{\rm rms} =
\ell_y/u_y^{\rm rms}=\ell_z/u_z^{\rm rms}$ together with the condition
$u_x^{\rm rms} \approx u_z^{\rm rms}$.

The flow in the chamber is inhomogeneous and anisotropic, so the assumption of identical turbulent time scales in all directions
and the assumption of $u_x^{\rm rms} \approx u_z^{\rm rms}$
represent approximations. The PIV measurements in our experiments provide two velocity components in the $YZ$ plane, while the third velocity component along the $X$ axis is not measured directly. Therefore, the resulting turbulent quantities should be regarded as estimates of the corresponding flow characteristics. However, our previous studies \cite{BEE04,EKRL26} have shown that these assumptions are valid for our experimental setups.

The turbulence is generated primarily by the forcing produced by the oscillating grids, and the large-scale shear also contributes to the turbulent kinetic energy. To visualize different flow regions, streamlines of the mean velocity field are superimposed on the spatial distributions of the turbulence characteristics.
As follows from the measurements, velocity fluctuations are stronger near the oscillating grids, and the turbulence intensity decreases with increasing distance $Y$ from the grids (see Figs.~\ref{Fig4} and~\ref{Fig7}--\ref{Fig8}). The turbulent velocity field generated by two oscillating grids is more symmetric in the horizontal direction and less anisotropic than that generated by one oscillating grid (see Figs.~\ref{Fig4} and~\ref{Fig6}). In most parts of the domain, the turbulent velocity exceeds the mean velocity (see Fig.~\ref{Fig5}).
Turbulent velocity varies from 2 to 17 cm/s for turbulence produced by one oscillating grid and from 19 to 32 cm/s for turbulence produced by two oscillating grids.

The integral turbulence length scales in turbulence produced by one oscillating grid are larger than those produced by two oscillating grids (see Figs.~\ref{Fig9}--\ref{Fig10}). The measured integral scales exhibit approximately linear dependencies, $\ell_i(Y) \propto Y$, in the left and right regions of the horizontal coordinate $Y$.
Similar scalings have been reported in earlier laboratory experiments in turbulent water flows produced by one oscillating grid \cite{turn68,turn73,tho75,hop76,kit97,san98,med01}, where the integral turbulence length scales are found to be increased linearly with the distance from the grid.

\subsection{Turbulent transport of particles}
\label{subsect4.2}

To investigate turbophoresis of inertial particles in inhomogeneous turbulence, we measure the spatial distributions of the normalized mean particle number density. The initial spatial distribution of the injected particles is nearly homogeneous and isotropic.
The spatial distributions of inertial and non-inertial particles are obtained from separate image sequences acquired
under identical flow conditions using the same optical configuration.

\begin{figure*}[t!]
\centering
\includegraphics[width=6.0cm]{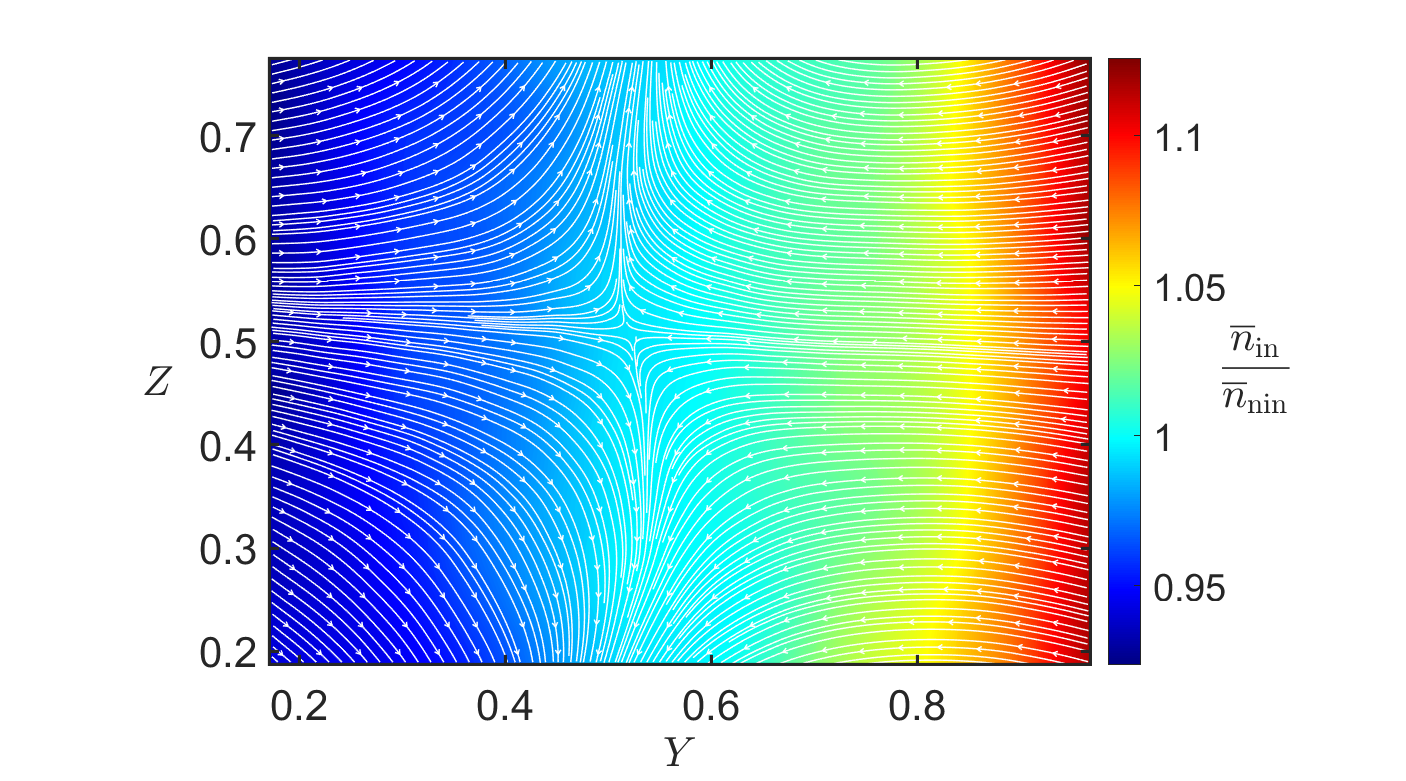}
\includegraphics[width=6.0cm]{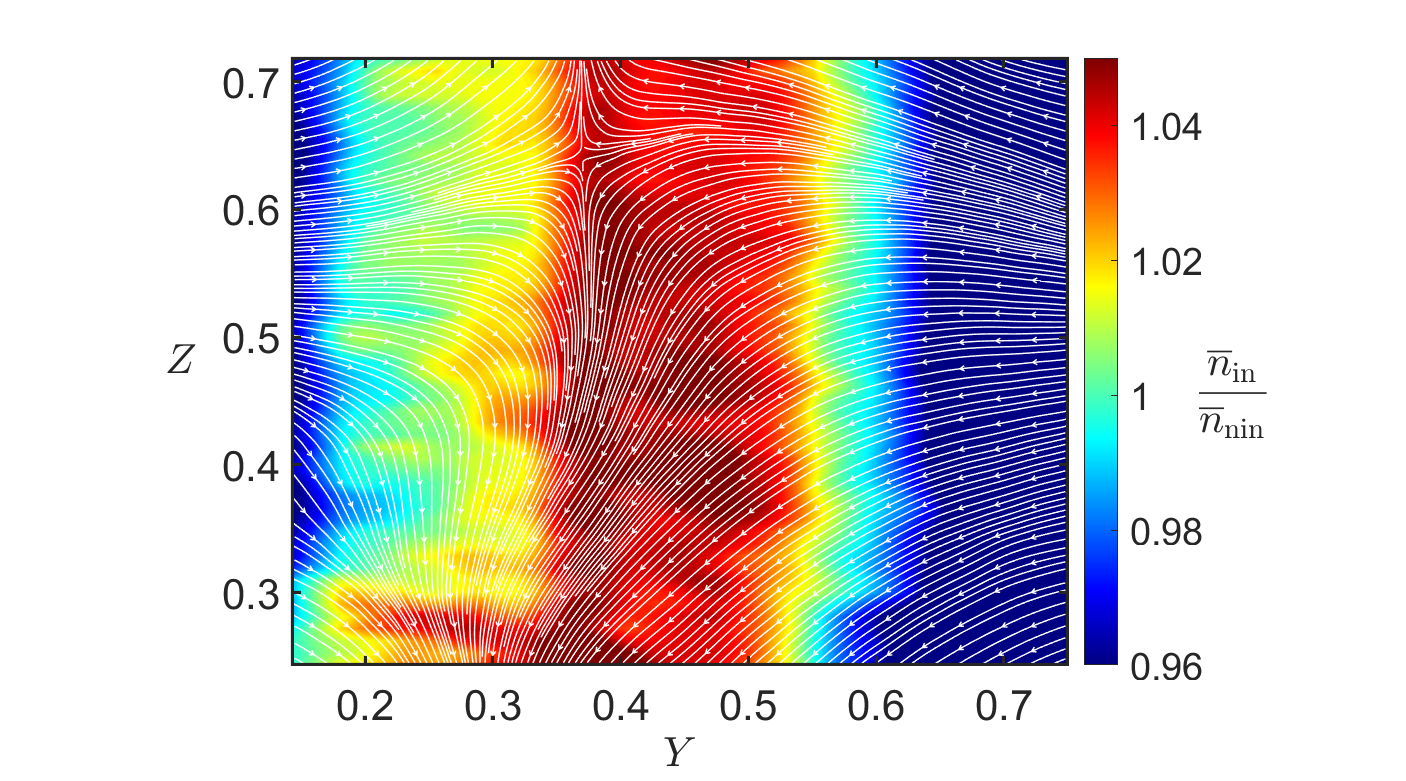}
\caption{\label{Fig12}
Spatial distributions of the normalized mean particle number density
$\meanN_{\rm in} / \meanN_{\rm nin}$
in the $YZ$ plane for turbulence generated by one oscillating grid (left panel) and two oscillating grids (right panel),
where $\meanN_{\rm in}$ is the mean number density of inertial particles
and $\meanN_{\rm nin}$ is the mean number density of noninertial particles.
Streamlines (white) of the mean velocity field $\meanU$ are superimposed on the particle distributions.
The coordinates $Y$ and $Z$ are normalized by $L_z=26$ cm.
}
\end{figure*}

\begin{figure*}[t!]
\centering
\includegraphics[width=6.0cm]{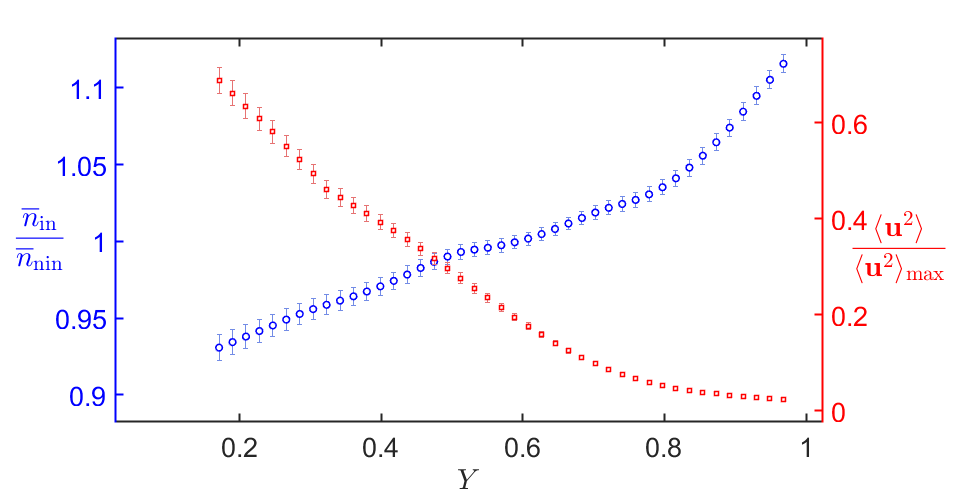}
\includegraphics[width=6.0cm]{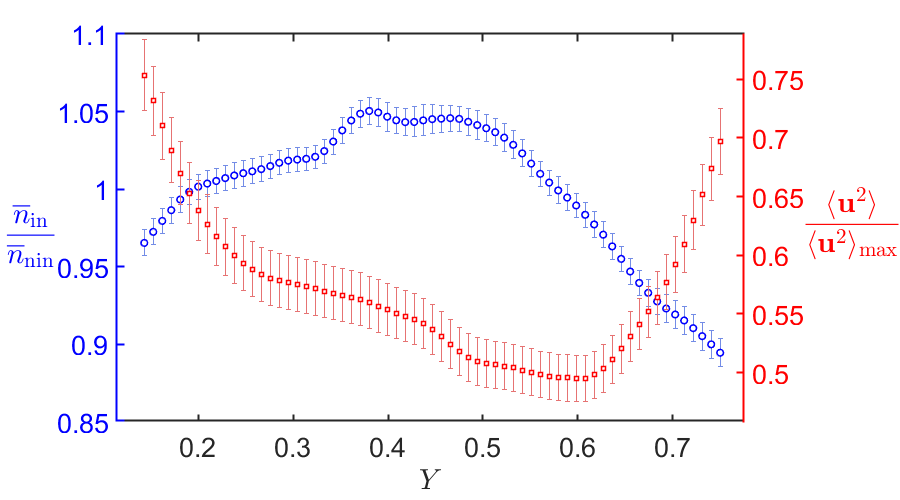}
\caption{\label{Fig13}
The $Y$-dependences of the normalized turbulence intensity
$\left\langle{\bm u}^2\right\rangle/ \left\langle{\bm u}^2\right\rangle_{\rm max}$ (red)
and the normalized mean particle number density
$\meanN_{\rm in} / \meanN_{\rm nin}$ (blue)
for turbulence generated by one oscillating grid (left panel) and two oscillating grids (right panel),
where $\meanN_{\rm in}$ is the mean number density of inertial particles
and $\meanN_{\rm nin}$ is the mean number density of noninertial particles.
The coordinate $Y$ is normalized by $L_z=26$ cm.
The error bars indicate propagated 95 \% statistical confidence intervals.
}
\end{figure*}

Figure~\ref{Fig12} shows the spatial distributions of the normalized mean particle number density $
\meanN_{\rm in} / \meanN_{\rm nin}
$ in the $YZ$ plane, where $\meanN_{\rm in}$ is the mean number density of inertial particles
and $\meanN_{\rm nin}$ is the mean number density of noninertial particles.
To analyze turbophoretic particle accumulation, we also plot in Fig.~\ref{Fig13} the $Y$-dependencies
of the normalized turbulence intensity
$\left\langle{\bm u}^2\right\rangle/ \left\langle{\bm u}^2\right\rangle_{\rm max}$,
and the normalized mean particle number density $\meanN_{\rm in} / \meanN_{\rm nin}$ for turbulence
produced by one oscillating grid (left panel) and two oscillating grids (right panel).
Here the turbulence intensity $\left\langle{\bm u}^2\right\rangle$ is normalised by its maximum
value $\left\langle{\bm u}^2\right\rangle_{\rm max}$.
Figures~\ref{Fig12}--\ref{Fig13} demonstrate that inertial particles preferentially accumulate in regions of lower turbulence intensity.

\begin{figure*}[t!]
\centering
\includegraphics[width=6.0cm]{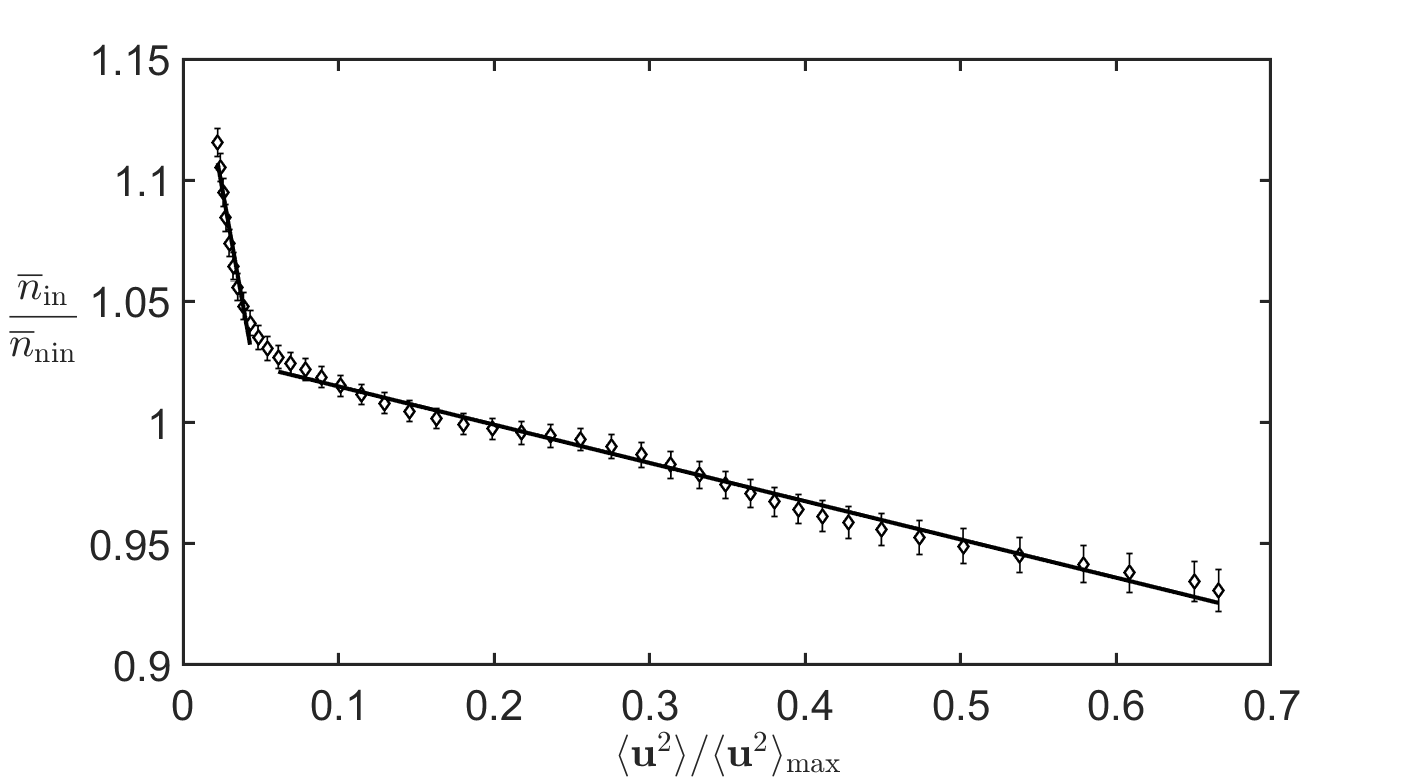}
\includegraphics[width=6.0cm]{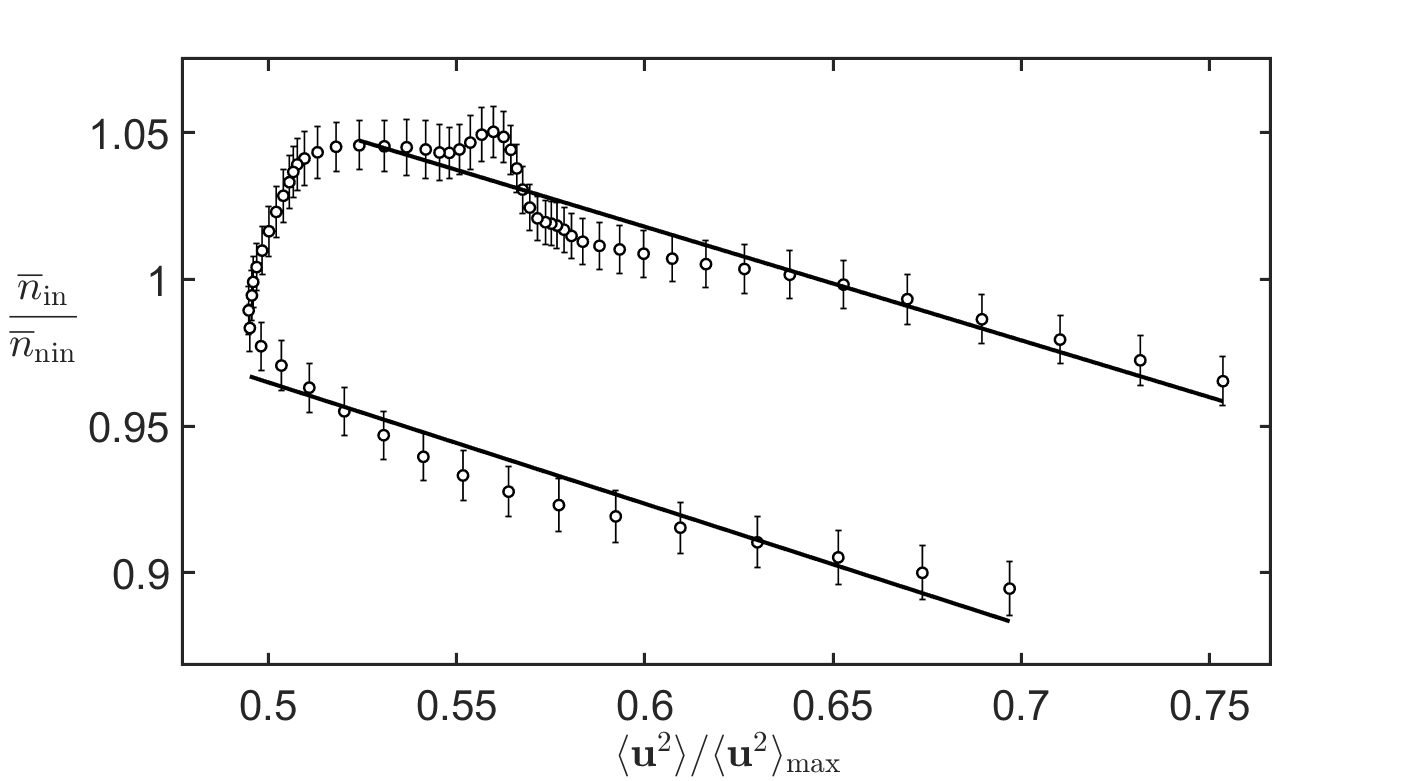}
\caption{\label{Fig14}
The normalized mean particle number density
$\meanN_{\rm in} / \meanN_{\rm nin}$
versus the normalized turbulence intensity
$\left\langle{\bm u}^2\right\rangle/ \left\langle{\bm u}^2\right\rangle_{\rm max}$
for turbulence generated by one oscillating grid (left panel) and two oscillating grids (right panel), where
$\left\langle{\bm u}^2\right\rangle_{\rm max}$
is the maximum turbulence intensity in the flow domain,
$\meanN_{\rm in}$ is the mean number density of inertial particles
and $\meanN_{\rm nin}$ is the mean number density of noninertial particles.
The error bars indicate propagated 95 \% statistical confidence intervals
of the normalized particle-number-density.
}
\end{figure*}

To compare the experimental results with the theoretical dependence given by Eq.~(\ref{G15}), Fig.~\ref{Fig14} shows the normalized mean particle number density $\meanN_{\rm in} / \meanN_{\rm nin}$ as the function of the normalized turbulence intensity $\left\langle{\bm u}^2\right\rangle/ \left\langle{\bm u}^2\right\rangle_{\rm max}$. The spatial distribution of noninertial particles remains nearly uniform throughout the domain, whereas inertial particles preferentially accumulate in regions of lower turbulence intensity. As a result,
the ratio $\meanN_{\rm in} / \meanN_{\rm nin}$ is smaller than 1 in most parts of the flow domain, except for the regions
where inertial particles are accumulated. These regions are located near the minima of the normalized turbulence intensity
$\left\langle{\bm u}^2\right\rangle/ \left\langle{\bm u}^2\right\rangle_{\rm max}$ (see Fig.~\ref{Fig14}).
This result is consistent with the theoretical estimate given by Eq.~(\ref{G15}).

In the two-grid configuration, the turbulence forcing is approximately symmetric with respect to the central region between the grids.
Therefore, the $Y$-profiles in Fig.~\ref{Fig14} (right panel) form two spatial branches corresponding to the two sides of the measurement region.
The solid lines are branchwise trend lines and are not theoretical fits.
They are shown only to indicate that both branches exhibit the same qualitative trend:
the normalized inertial-particle concentration increases as the normalized turbulence intensity decreases.

In Fig.~\ref{Fig14} we determine also the slopes of the trend lines (the solid lines shown in Fig.~\ref{Fig14})
which quantify the sensitivity of the normalized inertial-particle concentration
to changes in the normalized turbulence intensity.
They should not be interpreted as spatial gradients directly, but rather considered as a measure of how strongly the normalized particle concentration changes when the turbulence intensity varies.
For the one-grid configuration, the main branch gives a relatively moderate slope $(\approx -0.15$), while the localized branch at very low turbulence intensity gives a steeper slope $(\approx -3.17$). This indicates that the strongest change in the normalized particle concentration occurs in the region where the turbulence intensity is already low. This is consistent with the observed turbophoretic accumulation, since inertial particles preferentially accumulate in regions of reduced turbulence intensity.

For the two-grid configuration, the two branches have similar slopes
$(\approx -0.38$ for the upper line) and $(\approx -0.42$ for the lower line).
This supports the interpretation that the two branches do not represent separate physical mechanisms,
but instead they correspond to the decay of turbulence intensity away from the two oscillating grids.
Each branch represents one side of the approximately symmetric two-grid configuration,
and both branches show the same qualitative trend: the normalized inertial-particle concentration
increases as the turbulence intensity decreases.
The small deviation from the trend at the upper branch is related to the stagnation point of the mean flow,
but nevertheless it does not affect the whole trend.

The error bars shown in Figs.~\ref{Fig13}--\ref{Fig14} indicate propagated 95 \% statistical confidence intervals
of the normalized particle-number-density ratio, $\meanN_{\rm in} / \meanN_{\rm nin}$.
These uncertainties have been calculated directly from the data sets
for inertial and noninertial particles.
For each interrogation window, the standard error of the frame-averaged scattered-light intensity has been
evaluated separately for the inertial and noninertial particle measurements,
and the uncertainty of the ratio $\meanN_{\rm in} / \meanN_{\rm nin}$ has been obtained using first-order error propagation.
We distinguish between this conservative overall estimate
and the pointwise statistical confidence intervals shown in Figs.~\ref{Fig13}--\ref{Fig14}.

\section{Discussion and conclusions}
\label{sect5}

In the present experimental study, we have investigated turbophoretic transport of inertial particles in inhomogeneous non-stratified turbulence in airflow. Turbophoresis is caused by the combined effect of particle inertia and turbulence inhomogeneity. The competition between the effective turbophoretic velocity and turbulent diffusion can lead to the formation of large-scale concentrations of inertial particles in regions of minimum turbulence intensity.
In the experiments, the velocity field is measured using Particle Image Velocimetry, and the spatial distribution of inertial particles is obtained using image processing techniques.
To isolate the effect of turbophoresis, the particle number density measured for inertial particles is normalized pointwise by the corresponding distribution obtained for noninertial particles under identical flow conditions.

We examine the behavior of inertial particles in two distinct turbulent-flow configurations produced
by one and two oscillating grids.
Measurements of the velocity field show that these flow configurations have different spatial structure of turbulence intensity, anisotropy,
integral length scales, characteristic turbulent time, and Reynolds-number distributions.
Within the parameter range of the present experiments, both one-grid and two-grid configurations show the same qualitative behavior:
inertial particles preferentially accumulate in regions of reduced turbulence intensity, consistent with the theoretical estimate given by Eq.~(\ref{G15}).
In particular, the negative slopes of the solid trend lines shown in Fig.~\ref{Fig14}, indicate that
the normalized inertial-particle concentration increases as the normalized turbulence intensity decreases.
In the two-grid configuration, the two similar slopes correspond to the two spatial branches associated
with the decay of turbulence intensity away from each oscillating grid.
The larger apparent slope observed in the low-intensity branch of the one-grid configuration indicates a stronger local sensitivity of particle accumulation to changes in turbulence intensity in regions where the turbulence intensity is already low.

Although the mean fluid velocity field  has strong vertical structure (see Fig.~\ref{Fig2}),
these experiments have shown that the dominant effect of large-scale accumulation
of inertial particles occurs in the horizontal direction.
This indicates that the large-scale accumulation of inertial particles is mainly associated
with the horizontal inhomogeneity of turbulence rather than solely with the mean fluid velocity.

There might be an additional mechanism of particle accumulation
near the mean flow stagnation regions.
However, as can be seen in Fig.~\ref{Fig12} (left panel, corresponding to the one-grid configuration), the
regions where inertial particles accumulate  (at $Y$ from $0.83 L_z$ to  $0.9 L_z$)
are located far from the regions with the mean flow stagnation point (at $Y=0.55  L_z$).
As follows from Fig.~\ref{Fig12} (right panel, corresponding to the two-grid configuration), the
regions where inertial particles accumulate (at $Y$ from $0.33 L_z$ to  $0.53 L_z$ and all $Z$)
are different from the regions with the mean flow stagnation points (at $Y=0.36  L_z$ and $Z=0.65  L_z$).
In addition, as follows from our experiments, the spatial distribution of inertial particles is nonuniform.
This implies that the accumulation of inertial particles is
primarily driven by inhomogeneous turbulence rather than by the mean flow stagnation point.
The pointwise normalization of the inertial-particle number density by the corresponding noninertial-particle number density
reduces common mean-flow and stagnation-related effects.

The observed particle accumulation is primarily a bulk effect due to turbophoresis  caused by large-scale gradients of turbulence intensity rather than by boundary effects. This conclusion is supported by the measured spatial distributions of particle number density (see Figs.~\ref{Fig13}--\ref{Fig14}), which do not exhibit localization near the boundaries but instead follow the large-scale variations of turbulence intensity.
We do not use the present measurements to infer a quantitative value of
the turbophoretic coefficient $\kappa_{\rm turboph}$.
Such an estimate would require a fully three-dimensional characterization of the anisotropic inhomogeneous turbulence
and an independent determination of the turbulent diffusion tensor.
The comparison with previous theory and DNS is therefore made at the level of the observed spatial trend.
The normalized inertial-particle concentration increases in regions of reduced turbulence intensity,
consistent with the turbophoretic trend observed in DNS of inhomogeneously forced turbulence
and with the general physical picture of turbophoretic transport.

Although inertial particles experience gravitational settling, its influence on the measured particle accumulation remains limited under the present experimental conditions. The terminal fall velocity for particles with diameter $10\,\mu$m is approximately $0.33$ cm/s, which is substantially smaller than both the typical mean and turbulent fluid velocities.
Gravitational settling acts in the vertical direction, whereas the observed accumulation
trend is mainly associated with horizontal turbulence-intensity gradients.

The mass-loading parameter for inertial particles $n \, m_p/\rho \approx 6 \times 10^{-4}$
used in the experiments is very small (see Table~2),
so particles cannot affect the fluid flow field.
Furthermore, particle injection has been stopped before the measurements,
and no continuous particle supply was maintained during data acquisition.
The observed preferential accumulation of inertial particles occurs predominantly
in the horizontal direction and correlates with regions of reduced turbulence intensity,
whereas gravitational settling acts in the vertical direction.

The normalized particle-density distributions remain statistically stable during the measurements and show good repeatability across independent experimental runs.
Therefore, the present results provide experimental evidence of large-scale turbophoretic particle accumulation in strongly inhomogeneous turbulence.
A broad parametric experimental study of turbophoresis is the subject of a separate future investigation.

\bigskip
\noindent
{\bf ACKNOWLEDGEMENTS}
\medskip

We are thankful to the referees for providing constructive comments
which improved our paper.

\bigskip
\noindent
{\bf DATA AVAILABILITY}
\medskip

The data that support the findings of this study are available from the corresponding author
upon reasonable request.

\bigskip
\noindent
{\bf AUTHOR DECLARATIONS}

\medskip
{\bf  Funding Declaration}

This research did not receive funding.

\medskip
{\bf  Conflict of Interest}

The authors have no conflicts of interest to declare that are relevant to the content of this article.

\medskip
{\bf  Author Contributions}

Authors have contributed equally to the work.

\end{document}